\documentclass[useAMS, usegraphicx, usenatbib]{mn2e}
\usepackage{aas_macros}
\usepackage{amsmath}
\usepackage{graphicx}
\usepackage{natbib}

\newcommand{\munit}{\,h^{-1}\,\mathrm{M}_{\sun}}

\newcommand{\dNdz}{\mathrm{d}N/\mathrm{d}z}
\newcommand{\fN}{f_\mathrm{N}}

\newcommand{\wps}{\bar{w}_{\mathrm{ps}}}
\newcommand{\wpsm}{\widetilde{\bar{w}}_{\mathrm{ps}}}
\newcommand{\wss}{\bar{w}_{\mathrm{ss}}}
\newcommand{\bp}{\bar{b}_{\mathrm{p}}}
\newcommand{\bs}{\bar{b}_{\mathrm{s}}}
\newcommand{\bL}{b_{\mathrm{L}}}
\newcommand{\Np}{N_{\mathrm{p}}}
\newcommand{\Npm}{\widetilde{N}_{\mathrm{p}}}
\newcommand{\zmax}{z_{\mathrm{max}}}
\newcommand{\Mlim}{M_{\mathrm{lim}}}
\setcitestyle{authoryear,round,comma,aysep={},yysep={,},notesep={}}
\title[Cross-correlation-based luminosity functions]{A cross-correlation-based estimate of the galaxy luminosity function}
\author[M. P. van Daalen and M. White]{Marcel P. van Daalen$^{1,2}$\thanks{E-mail: marcel@berkeley.edu} and Martin White$^{1}$\\\\
$^1$Department of Astronomy, Theoretical Astrophysics Center, and Lawrence Berkeley National Laboratory,\\
\phantom{$^1$}University of California, Berkeley, CA 94720, USA\\
$^2$Leiden Observatory, Leiden University, P.O. Box 9513, 2300 RA Leiden, The Netherlands}
\begin{document}
\pagerange{\pageref{firstpage}--\pageref{lastpage}} \pubyear{2017}
\maketitle
\label{firstpage}
\begin{abstract}
We extend existing methods for using cross-correlations to derive redshift distributions for photometric galaxies, without using photometric redshifts. The model presented in this paper simultaneously yields highly accurate and unbiased redshift distributions and, for the first time, redshift-dependent luminosity functions, using only clustering information and the apparent magnitudes of the galaxies as input. In contrast to many existing techniques for recovering unbiased redshift distributions, the output of our method is not degenerate with the galaxy bias $b(z)$, which is achieved by modelling the shape of the luminosity bias. We successfully apply our method to a mock galaxy survey and discuss improvements to be made before applying our model to real data.
\end{abstract}

\section{Introduction}
\label{seq:introduction}
Current and planned large galaxy surveys are bringing in enormous amounts of photometric data. Spectroscopic follow-up for even a tenth of these sources is infeasible, and so many techniques have been developed to derive valuable redshift information indirectly for the vast majority of observed galaxies. Classically, estimating redshifts or redshift distributions has been performed using photometry in combination with a library of SEDs and/or spectroscopic sources to train the algorithms used, yielding a redshift (or redshift probability distribution) for each galaxy. However, these methods are not generally designed to yield unbiased redshift distributions, as they rely on the galaxies used in the training set to be representative of and similarly distributed to the overall population. Because of this, the accuracy of photometric redshifts (or photo-$z$s) can depend strongly on e.g.\ the magnitude, redshift and type of a galaxy, and the filters used \citep[e.g.][]{Cunha2009,Bezanson2016}.

While evolved methods exist that counter these problems \citep[e.g.][]{Lima2008}, one can also choose to avoid photometric redshifts altogether. One such way is to obtain redshift distributions for photometric galaxies statistically by examining how strongly they cluster with sources that have a known redshift. Even if these spectroscopic sources are a biased subset with a very different redshift distribution, they should still trace the same large-scale structure as the overall galaxy population. This means that it is statistically likely for galaxies to be at the same redshift as other sources they cluster strongly with, i.e.\ if two galaxies are close on the sky then they are more likely to be close along the line of sight. Techniques exploiting clustering to obtain independent redshift information have been applied for a number of years now, to improve and/or characterize the errors of a photo-$z$ catalogue \citep[e.g.][]{Padmanabhan2007,Newman2008,Erben2009,Kovac2010,Benjamin2010,QuadriWilliams2010,Choi2016}, reconstruct the density field \citep[e.g.][]{JascheWandelt2012,Malavasi2016,Cucciati2016} and to derive redshift distributions from clustering directly \citep[e.g.][]{MatthewsNewman2010,Schulz2010,McQuinnWhite2013,Menard2013,Morrison2017}. However, since this method is necessarily statistical we lose information on the properties of the galaxies in each redshift bin \citep[although recently efforts have been made to introduce a dependence on colour, see][]{Rahman2016}. Additionally, the resulting distribution is often degenerate with the unknown redshift-dependent bias of the photometric sample, $b(z)$, which has to be removed in some way before the outcomes can be used \citep[e.g.][]{Schmidt2013}.

Inspired by \citet{ShethRossi2010}, we extend existing methods to find the number density of galaxies in not only bins of redshift, $z$, but also apparent magnitude, $m$. By simultaneously fitting for both distributions, luminosity functions in terms of absolute magnitude, $M$, can be extracted at different redshifts. This has great potential, as the luminosity function is a key observable of the galaxy population that offers powerful constraints on models of galaxy evolution. Extensive cosmological volumes are needed to measure it accurately, particularly at the bright end where galaxies are rare. Large imaging surveys offer this, but their redshift uncertainties lead to uncertainties in the absolute magnitude of the galaxies. Spectroscopic surveys, on the other hand, have small redshift uncertainties but can probe far fewer galaxies. By cross-correlating these two types of survey while taking the observed brightness of the galaxies into account, we can derive luminosity functions for large volumes with smaller redshift uncertainties than would be possible otherwise. This method also allows us to break degeneracies in a new way: by assuming a simple model for just the luminosity dependence of the galaxy bias, the resulting redshift distributions and luminosity functions are independent of $b(z)$, and no bias removal is necessary.

We present our method for simultaneously deriving redshift distributions and luminosity functions from clustering data in \S\ref{sec:method}. As a test, we apply our model to a mock galaxy sample in \S\ref{sec:testing}. Finally, we summarize our results and discuss the possible limitations of our model when applied to real-world data in \S\ref{sec:discussion}.

\section{Methodology}
\label{sec:method}
The way in which we link the redshift distribution $\dNdz$ to the clustering signal can be viewed as a combination of the methods employed by \citet{Schulz2010} and \citet{Menard2013}, although we extend previous efforts by also estimating evolving luminosity functions for the photometric galaxies. Our approach is essentially to apply tomography to the luminosity function: the observed distribution of a sample of galaxies over apparent magnitude, $n(m)$, and the distributions of galaxies over redshift in bins of apparent magnitude, $n_\mathrm{m}(z)$, can be viewed as projections of the underlying luminosity function as a function of redshift, $\phi(M,z)$, and therefore used to reconstruct it. An added advantage of fitting for the redshift distributions and luminosity functions simultaneously is that it allows one to make optimal use of the information available in the survey -- for example, galaxies that appear bright are unlikely to be at high redshift.

In what follows, subscripts ``p'' denote the photometric sample for which we aim to derive a distribution in magnitude and redshift, while subscripts ``s'' denote the spectroscopic sample (which has a known redshift distribution).

\subsection{The cross-correlation signal}
\label{subsec:crosscorr}
The number of sample galaxies in apparent magnitude bin $m_\lambda$ and redshift bin $z_i$ is given by:
\begin{equation}
\label{eq:Np}
\Np(m_\lambda,z_i)=\int_{z_{i,\mathrm{min}}}^{z_{i,\mathrm{max}}}\int_{m_{\lambda,\mathrm{min}}}^{m_{\lambda,\mathrm{max}}}\frac{\mathrm{d}\Np}{\mathrm{d}m\,\mathrm{d}z}(m,z)\,\mathrm{d}m\,\mathrm{d}z,
\end{equation}
where "$i,\mathrm{min}$" and "$i,\mathrm{max}$" denote the edges of bin $i$. The parameter we wish to extract from the data is the fraction of sample galaxies in apparent magnitude bin $m_\lambda$ that reside in redshift bin $z_i$, given by:
\begin{equation}
\label{eq:fN}
\fN(m_\lambda,z_i)=\frac{\Np(m_\lambda,z_i)}{\Np(m_\lambda)},
\end{equation}
where $N_\mathrm{p}(m_\lambda)$ is the total number of galaxies in bin $m_\lambda$, given by:
\begin{equation}
\label{eq:Npm}
\Np(m_\lambda)=\sum_i \Np(m_\lambda,z_i).
\end{equation}
The $\Np(m_\lambda)$ of the data are known \textsl{a priori}, however we do not enforce the $\Np(m_\lambda)$ in our model -- which we will refer to as $\Npm(m_\lambda)$ -- to be identical to these. Rather, we interpret those in the data as being drawn from a Poisson distribution with means given by $\Npm(m_\lambda)$ (see \S\ref{subsec:Npmodel}).

As our signal we choose the integrated angular cross-correlation function of all photometric galaxies in apparent magnitude bin $m_\lambda$ with the spectroscopic galaxies in redshift bin $z_i$, $\wps(m_\lambda,z_i)$, given by:
\begin{equation}
\label{eq:wpsint}
\wps(m_\lambda,z_i)=\int_{\theta_\mathrm{min}}^{\theta_\mathrm{max}}w_\mathrm{ps}(m_\lambda,z_i,\theta)W(\theta)\,\mathrm{d}\theta,
\end{equation}
where $W(\theta)$ is a weight function. We follow \citet{Menard2013} in choosing $W(\theta)=\theta^{-1}$, and for the purposes of illustration choose $\theta_\mathrm{min}=0.02$ and $\theta_\mathrm{max}=10$ degrees.

We will refer to our model for $\wps(m_\lambda,z_i)$ as $\wpsm(m_\lambda,z_i)$. This quantity is related to the integrated angular correlation function between spectroscopic galaxies in redshift bin $z_i$ and those in redshift bin $z_j$, $\wss(z_i,z_j)$, through:
\begin{equation}
\label{eq:wps}
\wpsm(m_\lambda,z_i)=\sum_j \fN(m_\lambda,z_j) \frac{\bp(m_\lambda,z_j)}{\bs(z_j)} \wss(z_i,z_j),
\end{equation}
where $\bar{b}$ is the (linear) galaxy bias averaged over all scales $\theta$ between $\theta_\mathrm{min}$ and $\theta_\mathrm{max}$. Here we have used that the two samples trace the same underlying density field.

Both $\wps$ and $\wss$ can be directly calculated from the data (e.g.\ through pair counting), but the galaxy biases are \textsl{a priori} unknown. However, it is not unreasonable to assume that $\bp$ and $\bs$ evolve similarly with redshift at fixed luminosity, i.e.\ $\bp(m,z)=\bar{b}_\mathrm{p,0}\,\bL(m,z)f(z)$ and $\bs(z)=\bar{b}_\mathrm{s,0}\,f(z)$.\footnote{Alternatively, $\bs(z)$ could be estimated from the data (propagating the observational uncertainties) and $\bp(m,z)$ (or the ratio) could be modelled (e.g.\ as a polynomial in redshift).} Here $\bL$ is some function of luminosity -- assumed to be known, either independently or determined from the spectroscopic sample -- with no residual dependence on $m$ or $z$.

Next, we recognize that the redshift evolution $f(z)$ of the biases cancels out when taking the ratio, and absorb all constants in a new term. Then, in the limit of infinitely accurate measurements of $\wps(m_\lambda,z_i)$ and $\wss(z_i,z_j)$ we can derive $\fN$ simply by solving (for all $z_i$):
\begin{equation}
\label{eq:wpsconst}
\wpsm(m_\lambda,z_i)=\sum_j \fN'(m_\lambda,z_j) \wss(z_i,z_j),
\end{equation}
where $\fN'(m_\lambda,z_j)=K\bL(m_\lambda,z_j)\Np(m_\lambda,z_j)/\Np(m_\lambda)$ with $K$ an unknown constant and a parameter of the model. This set of equations can be written as $\mathbf{\wpsm}(m_\lambda)=\mathbf{X}\,\mathbf{\fN'}(m_\lambda)$, with $\mathbf{\wpsm}(m_\lambda)$ and $\mathbf{\fN'}(m_\lambda)$ vectors of length $n_\mathrm{z}$ and $\mathbf{X}$ a matrix of size $n_\mathrm{z} \times n_\mathrm{z}$, where $n_\mathrm{z}$ is the number of redshift bins. Hence, $X_{ij}=\wss(z_i,z_j)$. Note that we do not assume the often-used \citet{Limber1953} approximation, but allow for non-zero cross-correlations between redshift bins. Even though such cross-correlations are often serendipitous, they contain additional information on the large-scale density field and therefore can offer additional constraints. In \ref{subsubsec:limber}, we show how our results are affected if these cross-correlations are assumed to be zero.

For the purposes of illustration, we choose the following simple form for the luminosity bias \citep[motivated by e.g.][]{Benoist1996,Peacock2001,Norberg2001}:
\begin{equation}
\label{eq:bL}
\bL(m,z)=1+\frac{L(m,z)}{L'},
\end{equation}
where $L(m,z)$ is the luminosity of a galaxy of apparent magnitude $m$ at redshift $z$. We set $L'$ to be the luminosity of a galaxy with absolute magnitude $M'=-23.3$. Note that the normalization of the luminosity bias is indirectly controlled by the model parameter $K$. Since our model is agnostic about the redshift and therefore luminosity of each individual galaxy, we calculate the luminosity bias only once for each bin $(m_\lambda,z_i)$, assuming a naive relation between apparent and absolute magnitude (see \S\ref{subsec:Npmodel}).

At this point, we could solve the equations given by $\mathbf{\wpsm}(m_\lambda)=\mathbf{X}\,\mathbf{\fN'}(m_\lambda)$ for every $m_\lambda$ independently to find the corresponding galaxy redshift distributions. However, this disregards the information inherent in the apparent magnitudes of the galaxies. Since the clustering measurements have uncertainty (and since there may be degenerate solutions), this will likely lead to, for example, at least some galaxies with a very bright apparent magnitude being placed at high redshift -- corresponding to an unphysically high luminosity. Luminosity functions fitted to these results will therefore be extremely biased and unrealistic.

By fitting to the redshift distribution and the luminosity functions of the sample galaxies simultaneously, we avoid such biased outcomes. This requires us to explicitly model $\Np(m_\lambda,z_i)$.

For conciseness, we will use a subscript notation for binned quantities, i.e.\ $\Np{}_{,\lambda i}\equiv\Np(m_\lambda,z_i)$, where Greek subscripts always refer to the apparent magnitude bin and Latin subscripts to the redshift bin. Since we fit our model to all bins simultaneously, it is useful to think in terms of superindices $(\lambda i)=n_\mathrm{z}\lambda+i$. We will omit the parentheses where it does not lead to confusion.

\subsection{A model for $\Np(m,z)$}
\label{subsec:Npmodel}
\begin{figure*}
\begin{center}
\begin{tabular}{ccc}
\includegraphics[width=1.0\columnwidth, trim=27mm 8mm 13mm 8mm]{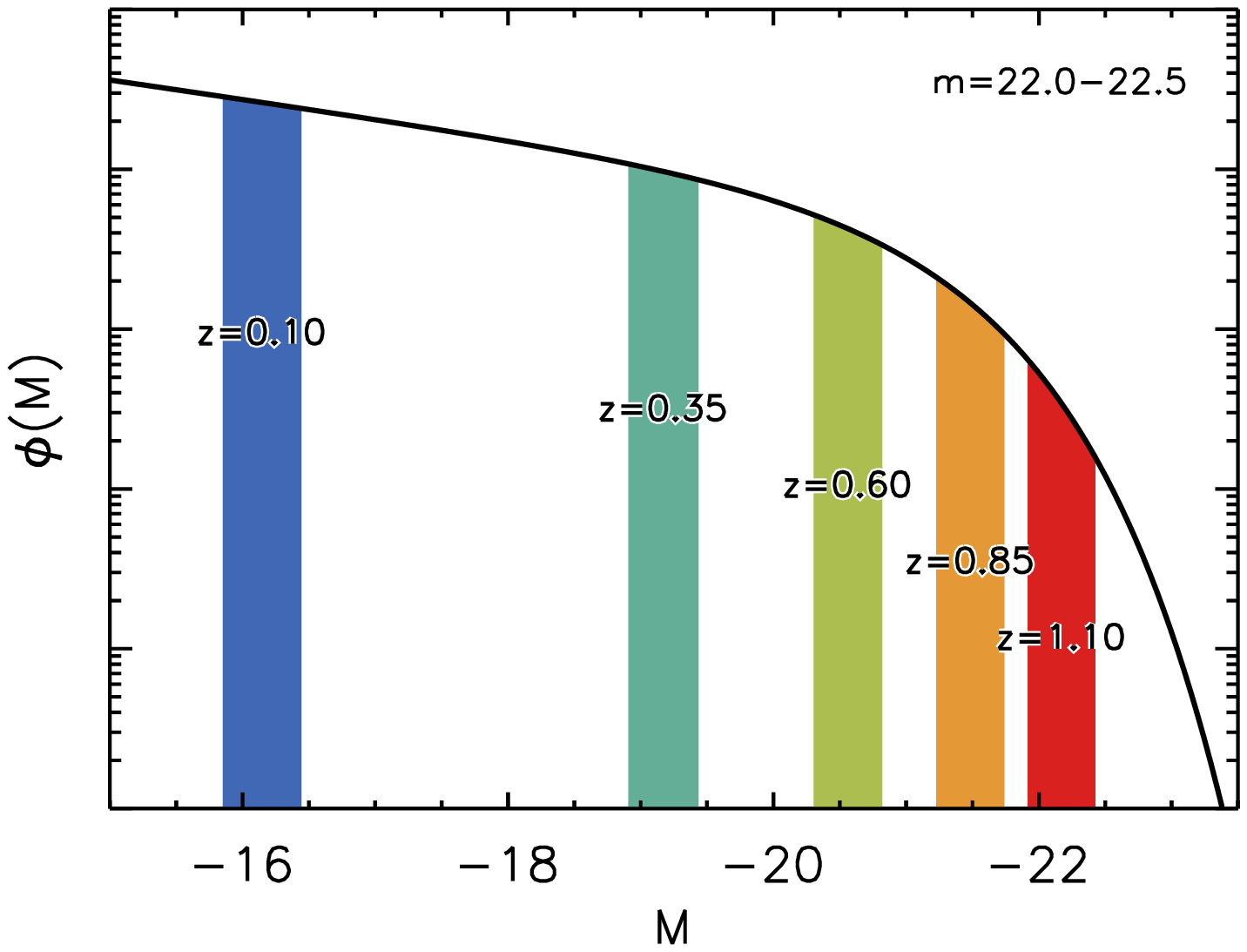} & &
\includegraphics[width=1.0\columnwidth, trim=37mm 8mm 3mm 8mm]{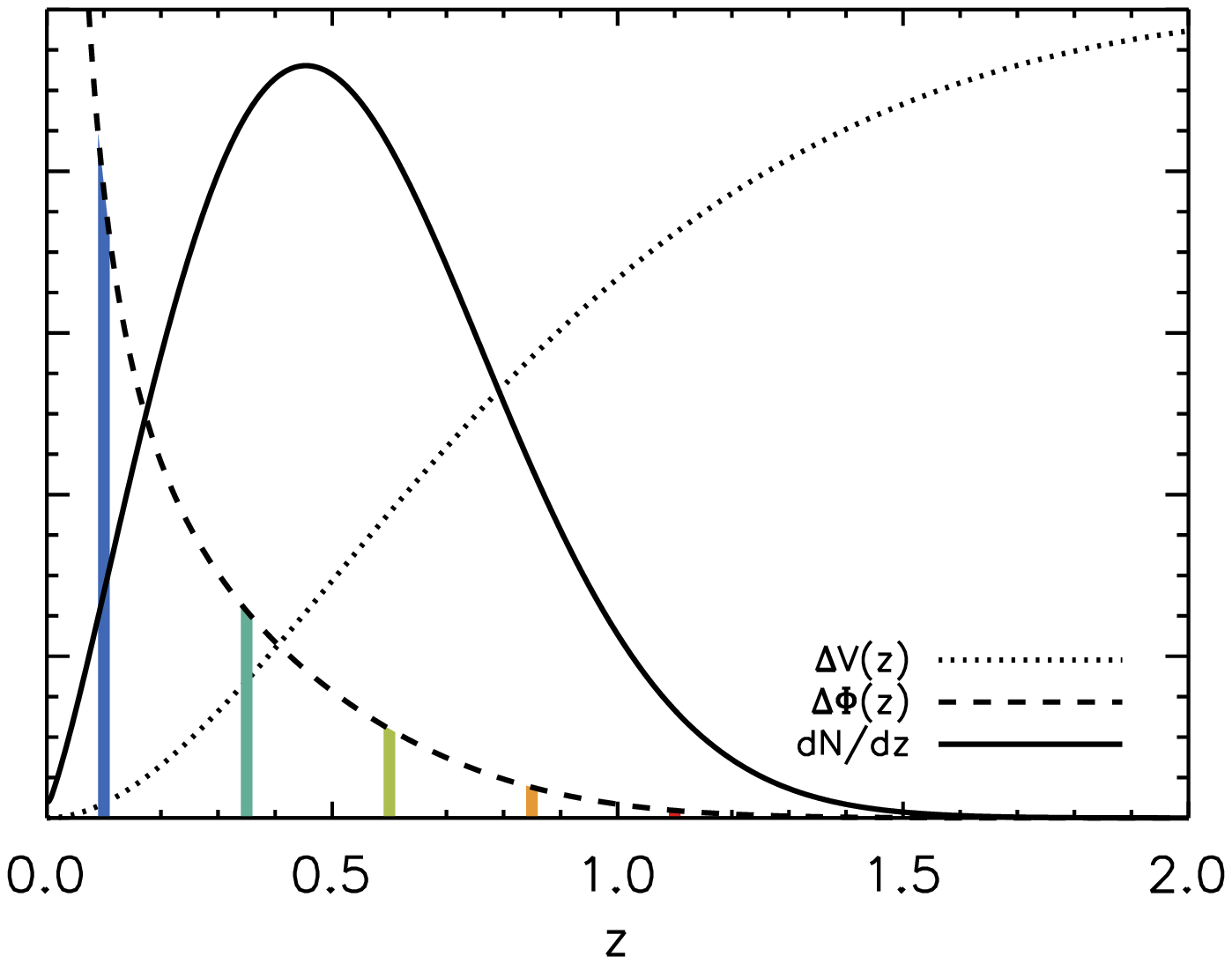}
\end{tabular}
\caption{The survey volume and the luminosity function combine to form the redshift distribution $\dNdz$. Shown here is an example for galaxies in an apparent magnitude bin $m=[22,22.5]$. \textit{Left:} A \citet{Schechter1976} luminosity function, with arbitrary normalization, as a function of absolute magnitude $M$. Here $(\alpha,M_*)=(-1.3,-21.1)$. Coloured regions show the absolute magnitudes corresponding to $m=[22,22.5]$ at redshifts $z=0.1$, $0.35$, $0.6$, $0.85$ and $1.1$ as indicated in the figure. In this example we assume the luminosity function is independent of redshift. \textit{Right:} Shown together here are the comoving volume added by each redshift slice, $\Delta V(z)$, the Schechter function shown on the left integrated over the relevant range in $M$, $\Delta\Phi(M)$, and the redshift distribution resulting from their product, $\dNdz=\Delta\Phi\Delta V$. The integral over one of the highlighted regions in the left-hand panel corresponds to the highlighted height of $\Delta\Phi(M)$ in the right-hand panel.}
\label{fig:dNdzex}
\end{center}
\end{figure*}

$\Np{}_{,\lambda i}$ is shaped by the luminosity function, which determines the number density of galaxies at apparent magnitude $m_\lambda$ and redshift $z_i$, and the survey volume at redshift $z_i$. Figure~\ref{fig:dNdzex} illustrates how these two quantities combine to form the redshift distribution of galaxies at fixed apparent magnitude. In this example we assume that both the luminosity function and the total number density of galaxies are constant with redshift. We consider galaxies in a fixed apparent magnitude bin, although the principle applies to any magnitude-limited survey. As the survey volume grows with redshift, the number of galaxies observed per unit redshift increases. However, galaxies with a fixed apparent magnitude correspond to increasingly more-luminous and more-rare galaxies, and so the number density decreases with redshift, first as a power law and then exponentially. The combined result of these two competing effects is a galaxy redshift distribution $\dNdz \propto \Delta\Phi\Delta V$ that increases as a power law before decreasing exponentially.\footnote{In the case of an evolving luminosity function, the integral over the sky and the luminosity function do not separate out as neatly as in this example, but the end result is similar. We do not make the assumption of a redshift-independent luminosity function beyond this example.}

Assuming a cosmology fixes the evolution of the survey volume. The shape of the luminosity function at each $z$ then fixes the redshift distribution. Conversely, knowing both the cosmology and the redshift distribution at several fixed apparent magnitudes gives us information on the shape of the luminosity function through cosmic time.

The comoving distance (for a flat $\Lambda$CDM universe) is given by:
\begin{equation}
\label{eq:codist}
d_\mathrm{c}(z)=\int_0^z \frac{c}{H_0\sqrt{\Omega_\mathrm{m,0}(1+z')^3+\Omega_{\Lambda,0}}}\mathrm{d}z',
\end{equation}
and hence the volume in redshift bin $i$ by:
\begin{eqnarray}
\nonumber
V_i \!\!\!\!\!&=&\!\!\!\!\! \int_A \int_{d_{i,\mathrm{min}}}^{d_{i,\mathrm{max}}}d_\mathrm{c}(z)^2\,\mathrm{d}d_\mathrm{c}(z)\,\mathrm{d}A\\
\label{eq:volume}
\!\!\!\!\!&=&\!\!\!\!\! f(A)\,4\pi\!\! \int_{z_{i,\mathrm{min}}}^{z_{i,\mathrm{max}}}\frac{d_\mathrm{c}(z)^2\,c}{H_0\sqrt{\Omega_\mathrm{m,0}(1+z)^3+\Omega_{\Lambda,0}}}\,\mathrm{d}z,
\end{eqnarray}
where $A$ is the area on the sky the survey covers and $f(A)$ is the fraction of the sky (in units of steradians) covered, and where the limits of integration $z_{i,\mathrm{min}}$ and $z_{i,\mathrm{max}}$ are the minimum and maximum redshift values respectively of redshift bin $i$.

For the purposes of illustration, we will assume the luminosity function is described well by a single Schechter function. We further assume that its parameters $\alpha$ (the low-luminosity power-law slope) and $M_*$ (the turn-over absolute magnitude) evolve linearly with redshift, that is $\alpha=\alpha_0+\alpha_\mathrm{e}\,z$ and $M_*=M_{*0}+M_\mathrm{*e}\,z$. The normalization of the luminosity function is allowed to evolve with redshift as well; specifically, we model it as the exponential of a 5th-order polynomial, as follows:
\begin{equation}
\label{eq:phistar}
\phi_*(z)=\exp\left(\sum_{j=0}^{j=5} \zeta_j\left[\frac{2z}{\zmax}-1\right]^j\right)\!,
\end{equation}
with $\zmax$ the maximum redshift considered and six free parameters $\zeta_j$.\footnote{The number of parameters used to fit $\phi_*(z)$ should be high enough to allow enough versatility, but much smaller than the number of redshift bins to ensure that it varies smoothly and that no (additional) degeneracies are introduced. We found that using a fifth-order polynomial strikes a nice balance. This particular form for $\phi_*(z)$ was chosen for numerical reasons (e.g.\ an easily calculable derivative).} Our luminosity function thus has 10 free parameters in total. We note that the luminosity function can be straightforwardly generalized to include e.g.\ additional Schechter terms or a more (or less) sophisticated redshift evolution.

To avoid divergence (and because there exists a minimum luminosity to what is considered a galaxy), we define a limiting galaxy absolute magnitude $\Mlim$. In this study we set $\Mlim=-16$, but we note that any sufficiently dim value of $\Mlim$ does not influence the outcome of the model. The (integrated) number density of galaxies in apparent magnitude bin $m_\lambda$ and redshift bin $z_i$ is then:
\begin{eqnarray}
\label{eq:phi}
\Phi_{\lambda i} \!\!\!\!\!&=&\!\!\!\!\! \frac{2}{5}\ln{(10)}\int_{z_{i,\mathrm{min}}}^{z_{i,\mathrm{max}}}\phi_*(z)\times\\
\nonumber
\!\!\!\!\!& &\!\!\!\!\! \!\!\!\!\!\!\!\!\!\!\int_{M_1}^{M_2} \frac{10^{\frac{2}{5}(M_*(z)-M(m,z))(\alpha(z)+1)}e^{-10^{\frac{2}{5}(M_*(z)-M(m,z))}}}{\Gamma\left(\alpha(z)+1,10^{\frac{2}{5}(M_*(z)-\Mlim)}\right)}\,\mathrm{d}M\,\mathrm{d}z,
\end{eqnarray}
where $M(m,z)$ is the absolute magnitude corresponding to a galaxy with apparent magnitude $m$ at redshift $z$.\footnote{In reality, the conversion from apparent to absolute magnitude would involve calculating a K-correction. Here we make the simplified assumption of a flat galaxy spectrum, in which case the K-correction is zero and $M(m,z)=m+5\left[1-\log_{10}(d_\mathrm{L}(z))\right]$, with $d_\mathrm{L}(z)$ the luminosity distance. We also ignore higher-order effects like lensing magnification. See Appendix~\ref{app:ignored} for a brief discussion.} Here we have defined the number density such that $\int_{-\infty}^{\Mlim}\frac{\mathrm{d}\Phi_{\lambda i}}{\mathrm{d}M_{\lambda i}}\,\mathrm{d}M\equiv\int_{-\infty}^{\Mlim}\phi_i(M)\,\mathrm{d}M=\int\phi_*(z)\,\mathrm{d}z_i$. The limits of integration for $M$ in equation~\eqref{eq:phi} are determined by the edges of the bins $m_\lambda$ and $z_i$, but the former are bounded above by $\Mlim$. That is, $M_1=\min\left\{M(m_\mathrm{\lambda,min};z),\Mlim\right\}$ and $M_2=\min\left\{M(m_\mathrm{\lambda,max};z),\Mlim\right\}$.

To account for evolution of the different functions within each redshift bin, we simultaneously integrate the volume and the luminosity function. The expected (Poisson mean) number of galaxies in apparent magnitude bin $m_\lambda$ and redshift bin $z_i$ is then given by:\footnote{We note here that we ignore the modulation of observed galaxy number densities due to lensing magnification, which causes a magnification bias.}
\begin{eqnarray}
\label{eq:Nlambdai_pre}
\nonumber
\Npm{}_{,\lambda i} \!\!\!\!\!&=&\!\!\!\!\! \int_{z_{i,\mathrm{min}}}^{z_{i,\mathrm{max}}}\int_{m_1}^{m_2}\frac{\mathrm{d}\phi_i(M)}{\mathrm{d}z_i}\frac{\mathrm{d}V}{\mathrm{d}z}\,\mathrm{d}m\,\mathrm{d}z\\
\!\!\!\!\!&=&\!\!\!\!\! \frac{2}{5}\ln{(10)}\,B\int_{z_{i,\mathrm{min}}}^{z_{i,\mathrm{max}}}\frac{d_\mathrm{c}(z)^2\,\phi_*(z)}{\sqrt{\Omega_\mathrm{m,0}(1+z)^3+\Omega_{\Lambda,0}}}\times\\
\nonumber
\!\!\!\!\!& &\!\!\!\!\! \!\!\!\!\!\!\!\!\!\!\!\!\!\! \int_{m_1}^{m_2}\frac{10^{\frac{2}{5}(M_*(z)-M(m,z))(\alpha(z)+1)}e^{-10^{\frac{2}{5}(M_*(z)-M(m,z))}}}{\Gamma\left(\alpha(z)+1,10^{\frac{2}{5}(M_*(z)-\Mlim)}\right)}\,\mathrm{d}m\,\mathrm{d}z,
\end{eqnarray}
where some of the constants have been absorbed into the constant $B$; specifically, $B=4\pi f(A)\,c/H_0$. We have switched the integral over $M$ to an integral over $m$, but similar to before, $m_1=\min\left\{m_\mathrm{\lambda,min};m(\Mlim,z)\right\}$ and $m_2=\min\left\{m_\mathrm{\lambda,max};m(\Mlim,z)\right\}$. The integral over apparent magnitude has an analytical solution, and so we can we can reduce the above expression for the Poisson mean to an integral over only the redshift bin $z_i$:
\begin{eqnarray}
\nonumber
\Npm{}_{,\lambda i} \!\!\!\!\!&=&\!\!\!\!\! B\int_{z_{i,\mathrm{min}}}^{z_{i,\mathrm{max}}}\frac{d_\mathrm{c}(z)^2\,\phi_*(z)}{\sqrt{\Omega_\mathrm{m,0}(1+z)^3+\Omega_{\Lambda,0}}}\times\\
\nonumber
\!\!\!\!\!& &\!\!\!\!\! \left[\Gamma\left(\alpha(z)+1,10^{\frac{2}{5}(M_*(z)-\Mlim)}\right)\right]^{-1}\times\\
\nonumber
\!\!\!\!\!& &\!\!\!\!\! \left[\Gamma\left(\alpha(z)+1,10^{\frac{2}{5}(M_*(z)-M(m_2,z))}\right)-\right.\\
\label{eq:Nlambdai}
\!\!\!\!\!& &\!\! \left.\Gamma\left(\alpha(z)+1,10^{\frac{2}{5}(M_*(z)-M(m_1,z))}\right)\right]\,\mathrm{d}z.
\end{eqnarray}
The total number of model galaxies in apparent magnitude bin $m_\lambda$ at any redshift is then:
\begin{eqnarray}
\nonumber
\Npm{}_{,\lambda} \!\!\!\!\!&=&\!\!\!\!\! \sum_i \Npm{}_{,\lambda i}\\
\nonumber
\!\!\!\!\!&=&\!\!\!\!\! B\int_{z_\mathrm{min}}^{z_\mathrm{max}}\frac{d_\mathrm{c}(z)^2\,\phi_*(z)}{\sqrt{\Omega_\mathrm{m,0}(1+z)^3+\Omega_{\Lambda,0}}}\times\\
\nonumber
\!\!\!\!\!& &\!\!\!\!\! \left[\Gamma\left(\alpha(z)+1,10^{\frac{2}{5}(M_*(z)-\Mlim)}\right)\right]^{-1}\times\\
\nonumber
\!\!\!\!\!& &\!\!\!\!\! \left[\Gamma\left(\alpha(z)+1,10^{\frac{2}{5}(M_*(z)-M(m_2,z))}\right)-\right.\\
\label{eq:Nlambda}
\!\!\!\!\!& &\!\! \left.\Gamma\left(\alpha(z)+1,10^{\frac{2}{5}(M_*(z)-M(m_1,z))}\right)\right]\,\mathrm{d}z.
\end{eqnarray}
Here $z_\mathrm{min}$ and $z_\mathrm{max}$ are the limits of the redshift range probed by the spectroscopic sample. These model estimates of the mean can be directly compared to the $\Np{}_{,\lambda}$ of the data as a measure of our model's accuracy for a given set of parameters.

\begin{figure*}
\begin{center}
\begin{tabular}{ccc}
\includegraphics[width=1.0\columnwidth, trim=22mm 8mm 8mm 8mm]{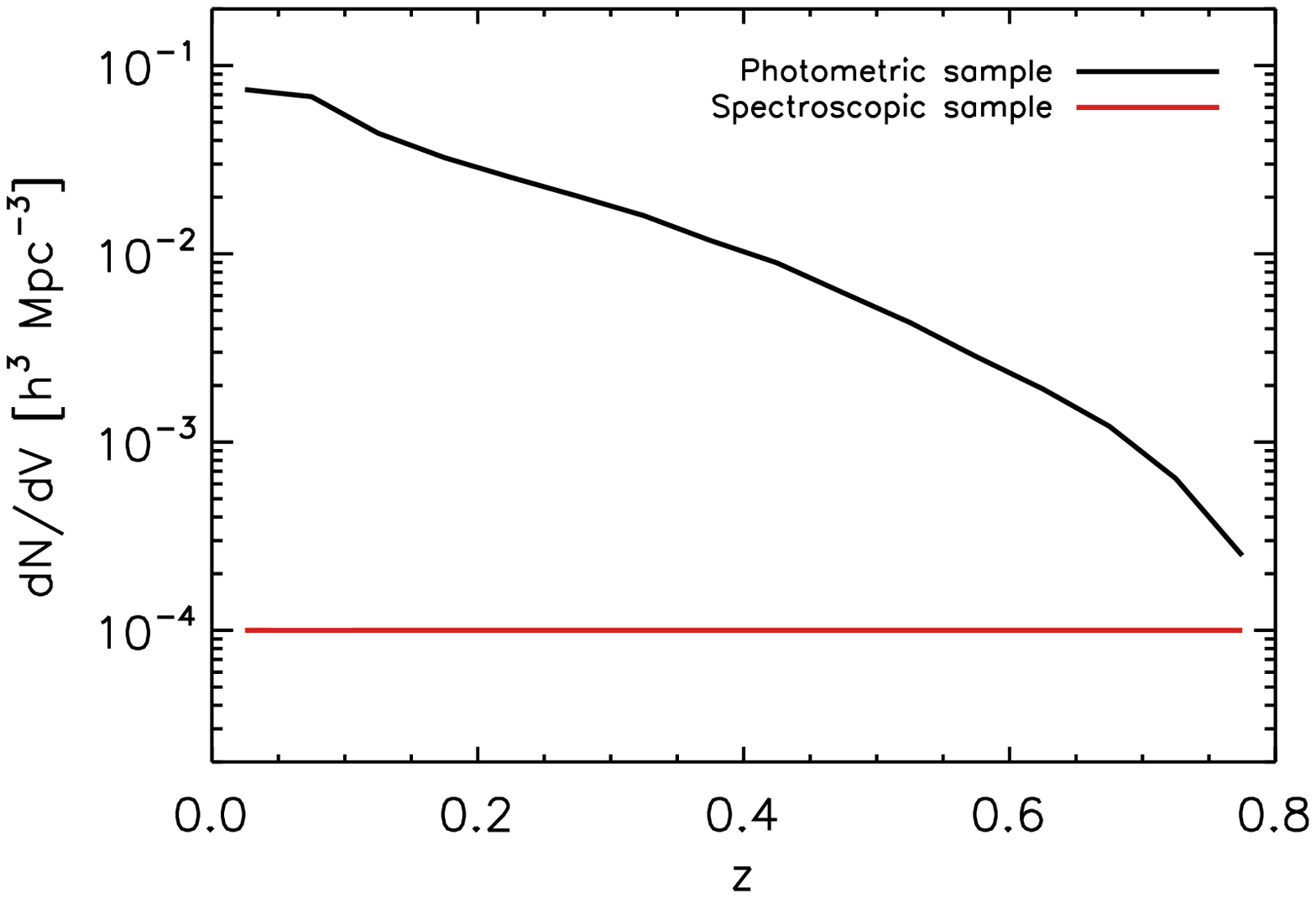} & &
\includegraphics[width=1.0\columnwidth, trim=22mm 8mm 8mm 8mm]{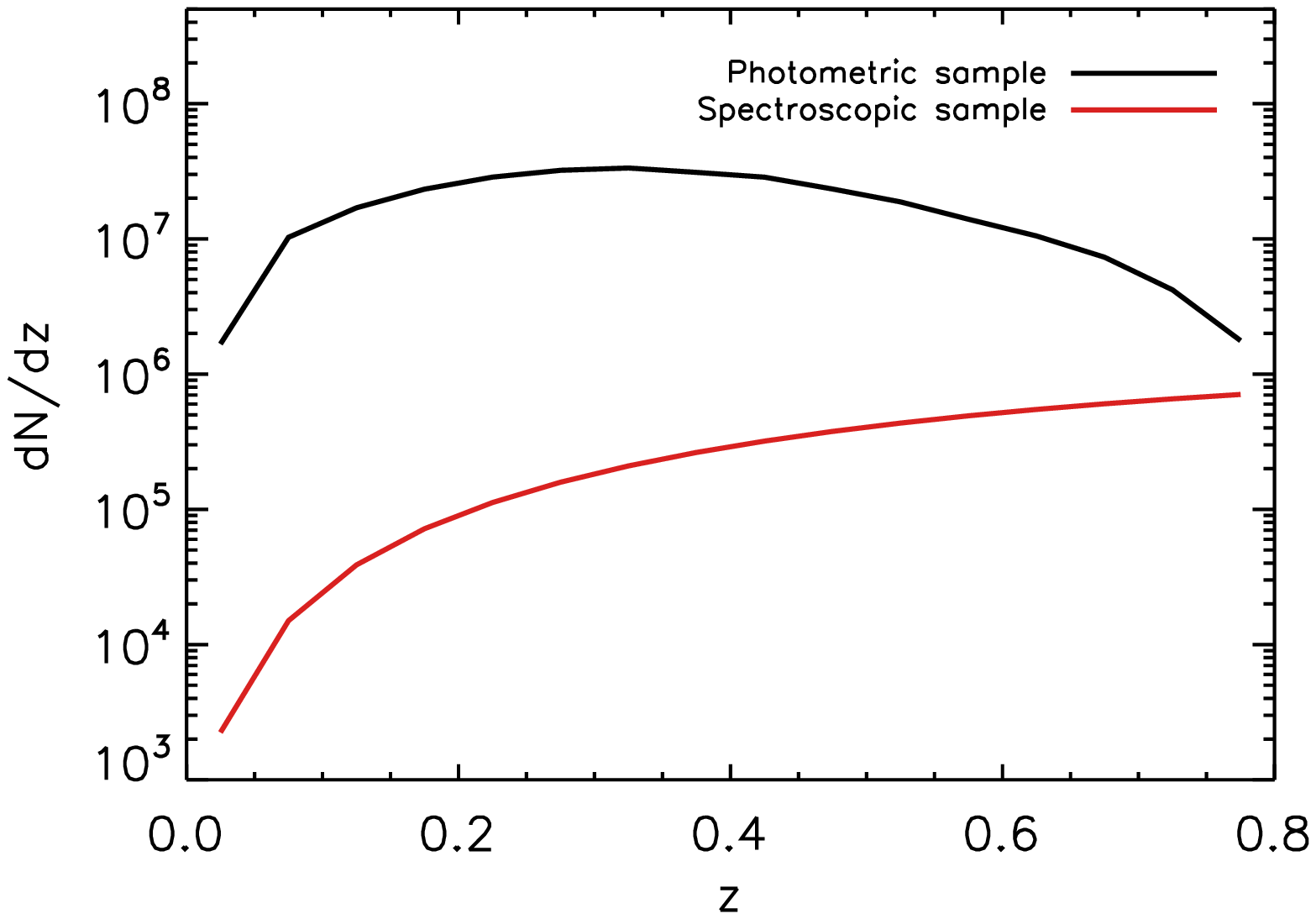}\\
\includegraphics[width=1.0\columnwidth, trim=22mm 8mm 8mm 8mm]{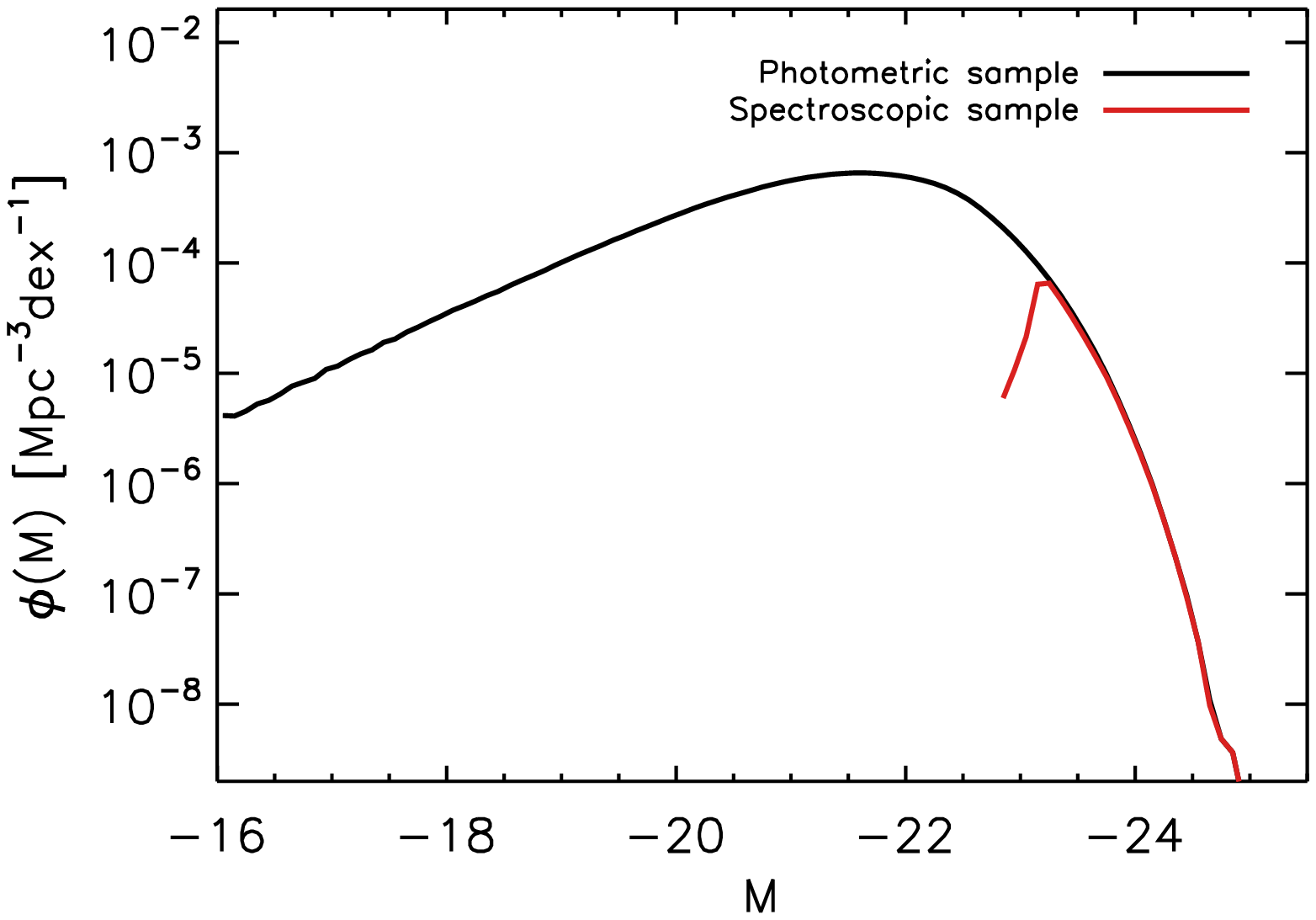} & &
\includegraphics[width=1.0\columnwidth, trim=22mm 8mm 8mm 8mm]{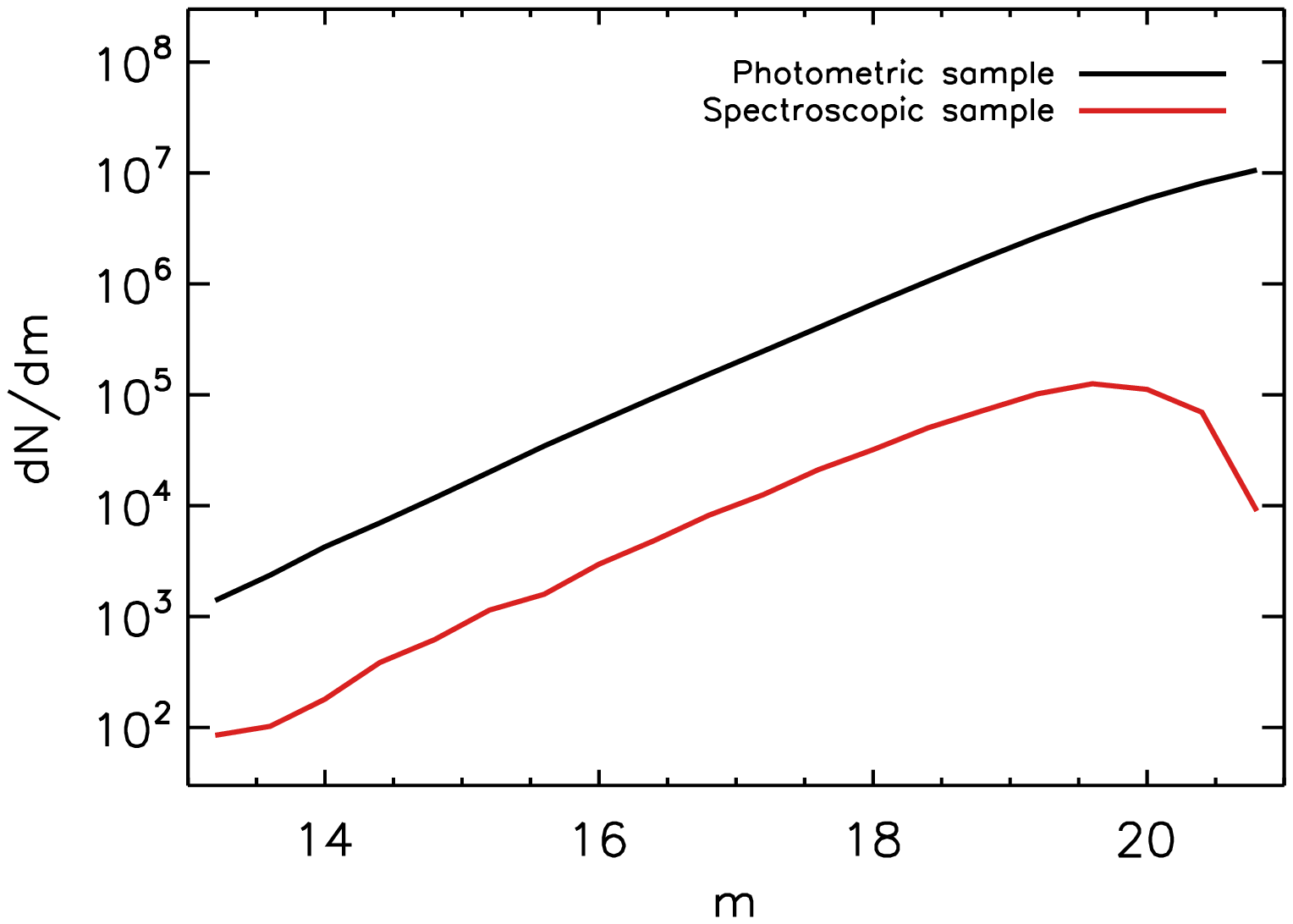}
\end{tabular}
\caption{A comparison of the distributions of our photometric (in black) and spectroscopic (in red) mock galaxy samples over cosmic (comoving) volume (top left), redshift (top right), absolute magnitude (bottom left) and apparent magnitude (bottom right). By construction, the spectroscopic sample has a spatial density of $10^{-4}\,(\mathrm{Mpc}/h)^{-3}$ over the entire redshift range and contains only the most luminous (star-forming) galaxies. Even though the spectroscopic sample is, realistically, a biased subset of the total galaxy population, it can still be used to derive accurate redshift distributions and luminosity functions for the photometric galaxies, as they trace the same large-scale structure and the clustering bias of the samples does not need to be known in our model.}
\label{fig:specvsphot}
\end{center}
\end{figure*}

\subsection{Fitting the model}
\label{subsec:modelfit}
Using 11 free parameters in total (1 parameter for the bias ratio, 6 for the normalization of the luminosity function and 4 for its shape parameters), our model predicts a distribution of galaxies in both absolute magnitude and redshift, and -- using the observed integrated autocorrelation of the spectroscopic sample $\wss{}_{,ij}$ -- the corresponding cross-correlation signal $\wpsm{}_{,\lambda i}$. The best-fit set of parameters is determined by comparing the model outcomes $\wpsm{}_{,\lambda i}$ and $\Npm{}_{,\lambda}$ to their observed counterparts. We fit for these two quantities simultaneously by minimizing:
\begin{equation}
\label{eq:chisq}
\chi^2=(\mathbf{\wps}-\mathbf{\wpsm})^\mathrm{T} C^{-1}(\mathbf{\wps}-\mathbf{\wpsm}) + R\sum_\lambda \frac{(\Np{}_{,\lambda}-\Npm{}_{,\lambda})^2}{\widetilde{\sigma}_{\lambda}^2},
\end{equation}
where $C$ is a joint covariance matrix combining different sources of uncertainty in both the data and the model (see Appendix~\ref{app:covar}), $R$ is a constant determining the relative weight of the two observables, and $\widetilde{\sigma}_{\lambda}^2$ is the variance of $\Npm{}_{,\lambda}$. Since $\Npm{}_{,\lambda}$ is a Poisson mean, $\widetilde{\sigma}_{\lambda}^2=\Npm{}_{,\lambda}$. The ideal value of $R$ is unknown, but it should be set such that $\Npm{}_{,\lambda}$ is not fit at the expense of $\wpsm{}_{,\lambda i}$, but rather used to break degeneracies in the clustering. In what follows, we set $R=n_\mathrm{z}$, so as to give the cross-correlation signal and galaxy number counts equal weight (after all, the former provides $n_\mathrm{z}\times n_\mathrm{m}$ data points while the latter only provides $n_\mathrm{m}$). When the Limber approximation is taken, we set $R=1$, since in this case the effective number of data points obtained from the clustering signal goes down by a factor of $n_\mathrm{z}$. Very similar results are obtained if we vary $R$ within a factor of $10$.

\section{Testing the model}
\label{sec:testing}

\subsection{Mock catalogues}
\label{subsec:mocks}
To test our model, we extract a mock galaxy survey from one of the publicly available Planck Millennium all-sky lightcones released with \citet{Henriques2015}.\footnote{Specifically, we use the catalogues ``cones.AllSky\_M05\_001'' and ``MRscPlanck1'' from the ``Henriques2015a'' part of the Millennium public database.} The semi-analytical model that forms the base for this lightcone is detailed in \citet{Henriques2015}, while information on how the lightcone was constructed and magnitudes were assigned can be found in \citet{Henriques2012}. In order to measure our model's performance, we have to know the luminosity function of the data. A potential mismatch in our final results may be due to either inaccuracies in the model or to the fact that a single Schechter function is not a perfect fit to the intrinsic luminosity function of the mock galaxies. In order to separate these effects, we reassign the absolute magnitude of each galaxy (in the $i$-band) so that it is consistent with an input luminosity function. This is done in redshift bins $0.01$ wide, and in such a way that the rank ordering of galaxies in brightness in each redshift bin is preserved (i.e.\ the $N$ brightest galaxies at each redshift before reassignment are still the $N$ brightest galaxies after reassignment, for every $N$). We choose the shape parameters of our Schechter function to be $\{\alpha_0,\alpha_\mathrm{e},M_{*0},M_\mathrm{*e}\}=\{-1.01,-0.15,-21.5,-0.8\}$ (see \S\ref{subsec:Npmodel}). These parameters were chosen in fair approximation of the intrinsic luminosity function of the galaxies in the lightcone. We do not change the redshifts, locations or number densities as a function of redshift of the galaxies, hence the normalization of the luminosity function and the clustering bias of the galaxies is still determined by the processes that formed them. The apparent magnitude of each galaxy is recalculated to match its new absolute magnitude, assuming again a naive relation between these and redshift (i.e.\ without K-corrections). We then make cuts in apparent magnitude and redshift, only keeping galaxies for which $m \le 21$ and $z \le 0.8$. Next, we arbitrarily select the region with right ascension within $[100,200]$ degrees and declination within $[10,50]$ degrees, equivalent to $3394\,\mathrm{deg}^2$ or about $8\%$ of the sky.

The galaxies that are left comprise our photometric galaxy sample. From it, we select a spectroscopic sample by selecting the $N$ brightest galaxies in each redshift bin with stellar masses $M_* \ge 10^{10}\munit$ and star formation rates $\dot{M}_* \ge 1\munit/\mathrm{yr}$, where $N$ is chosen such that the number density of spectroscopic galaxies is at most $10^{-4}\,(\mathrm{Mpc}/h)^{-3}$ at every redshift.\footnote{To clarify, after our pre-selection by stellar mass and star formation rate, we choose the $N$ galaxies at each redshift that are brightest in absolute magnitude. One might argue that a more natural choice is apparent magnitude; however, since we make our selection in separate redshift bins, the difference is minimal.} Note that for the purpose of demonstrating the effectiveness of our model, all that matters is that the spectroscopic sample is a small and highly biased subset of the total population, not that it is realistically selected. For the photometric sample we retain only the position on the sky and apparent magnitude. Finally, we take $n_\mathrm{z}=16$ redshift bins, $\Delta z=0.05$ wide, in the range $z=[0,0.8]$, and $n_\mathrm{m}=16$ apparent magnitude bins, $\Delta m=0.5$ wide, in the range $m=[13,21]$, and calculate the relevant \mbox{(cross-)correlation} functions and covariance matrices. In total, our photometric sample contains $14,\!280,\!584$ galaxies that fall in these ranges, and our spectroscopic sample contains $250,\!372$ sources.

\begin{figure*}
\begin{center}
\large \medskip\quad\quad\textbf{Fiducial model}\par\medskip
\begin{tabular}{ccc}
\includegraphics[width=1.0\columnwidth, trim=22mm 8mm 8mm 8mm]{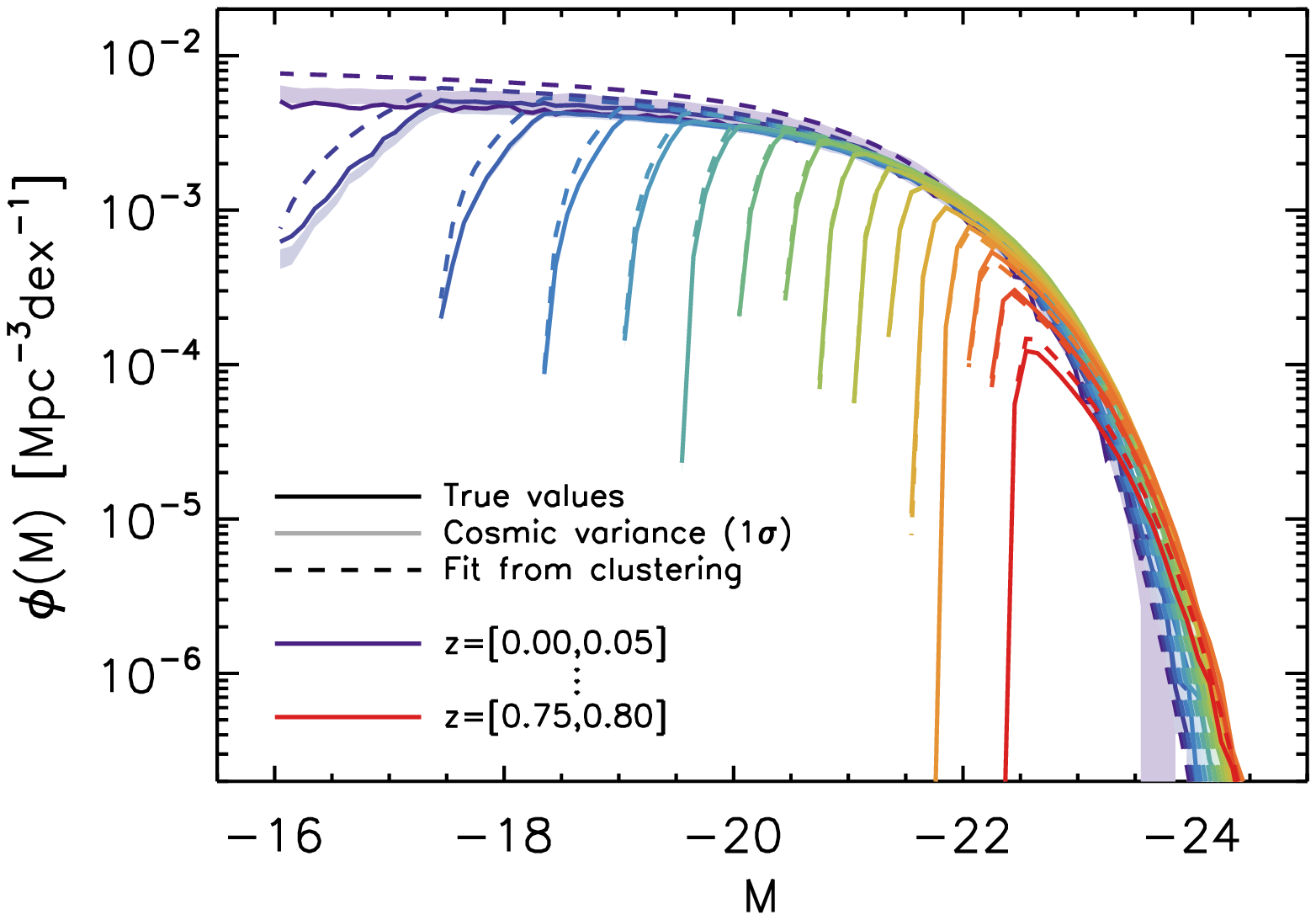} & &
\includegraphics[width=1.0\columnwidth, trim=22mm 8mm 8mm 8mm]{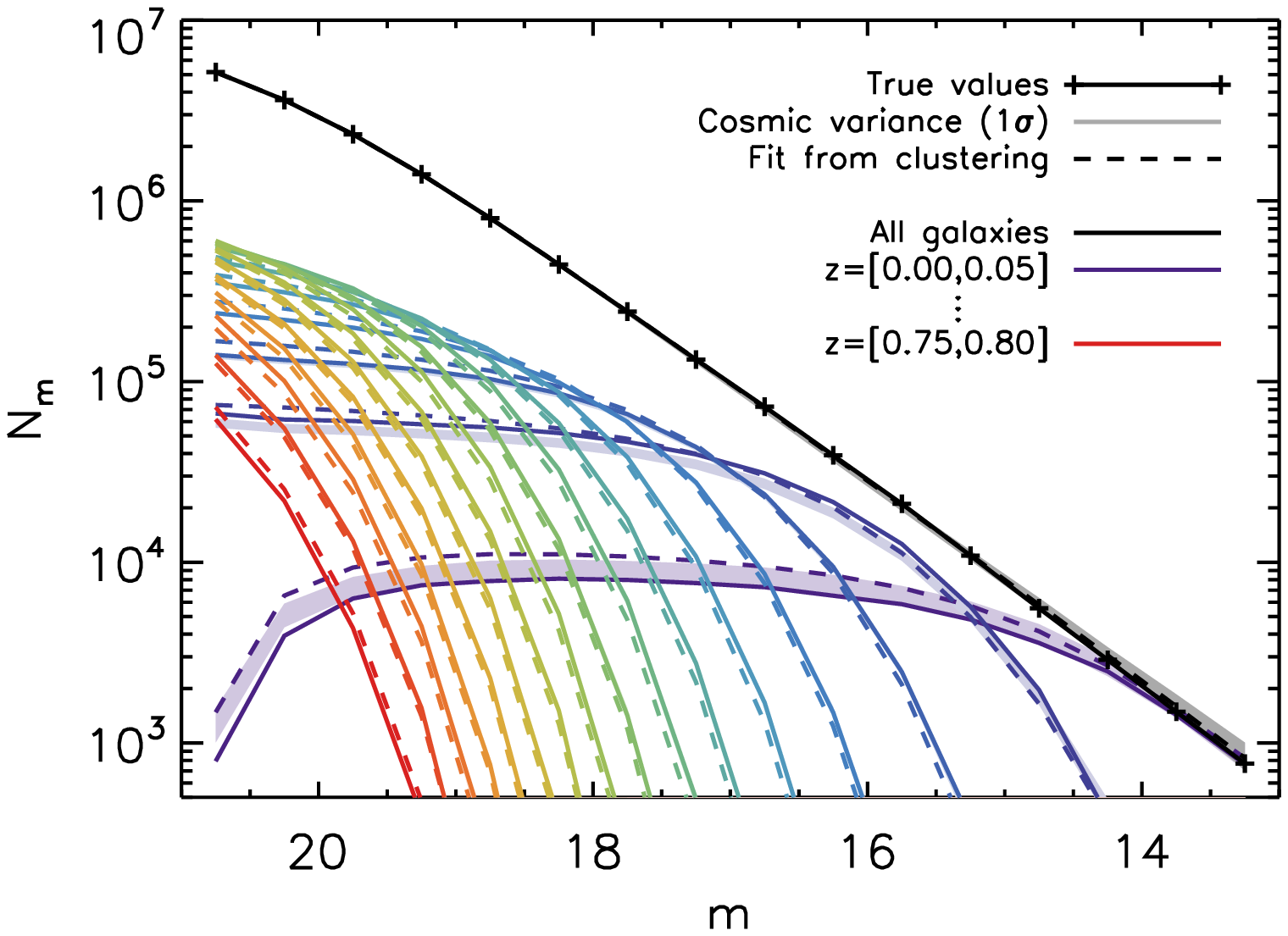}\\
\includegraphics[width=1.0\columnwidth, trim=22mm 8mm 8mm 8mm]{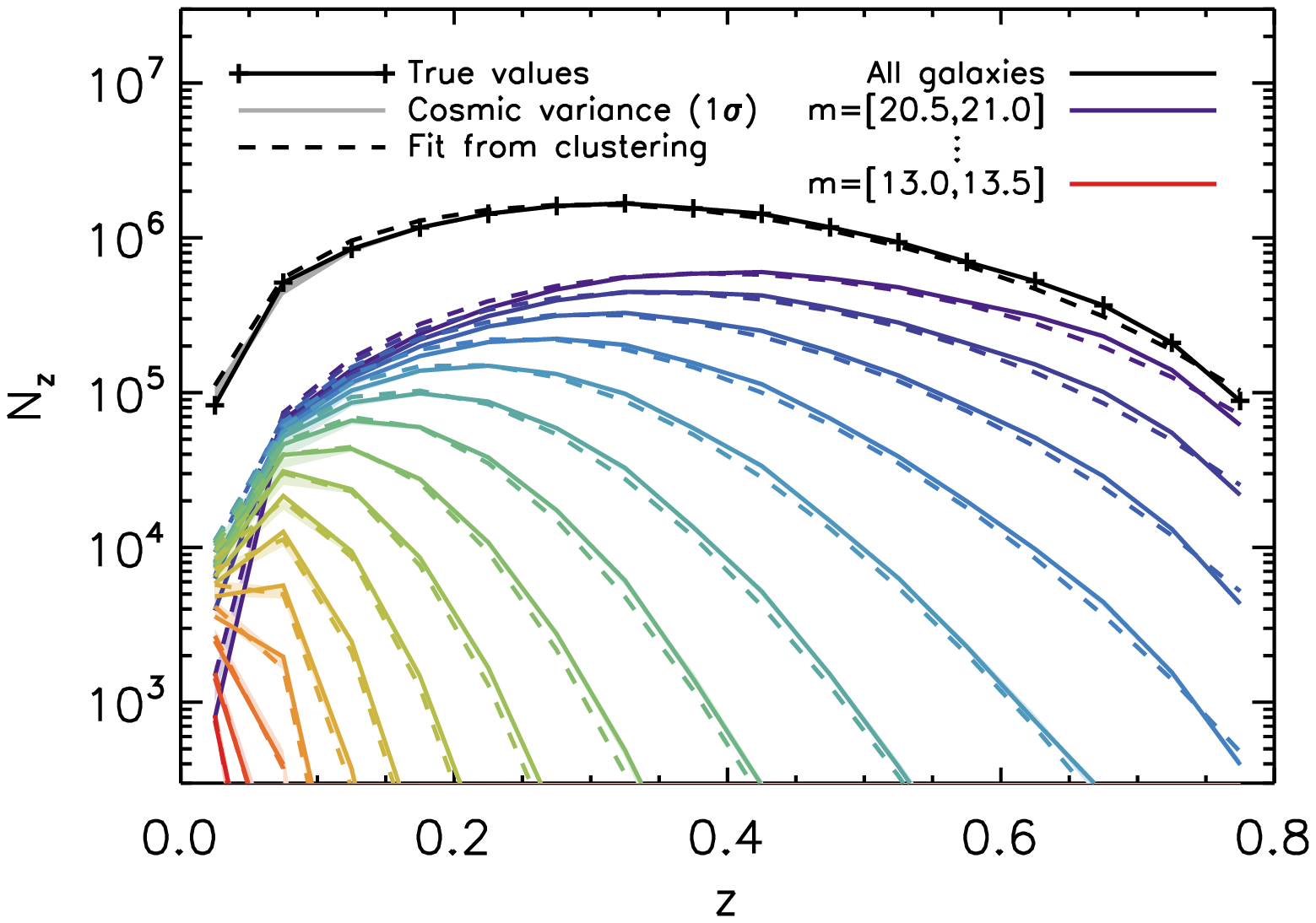} & &
\includegraphics[width=1.0\columnwidth, trim=22mm 8mm 8mm 8mm]{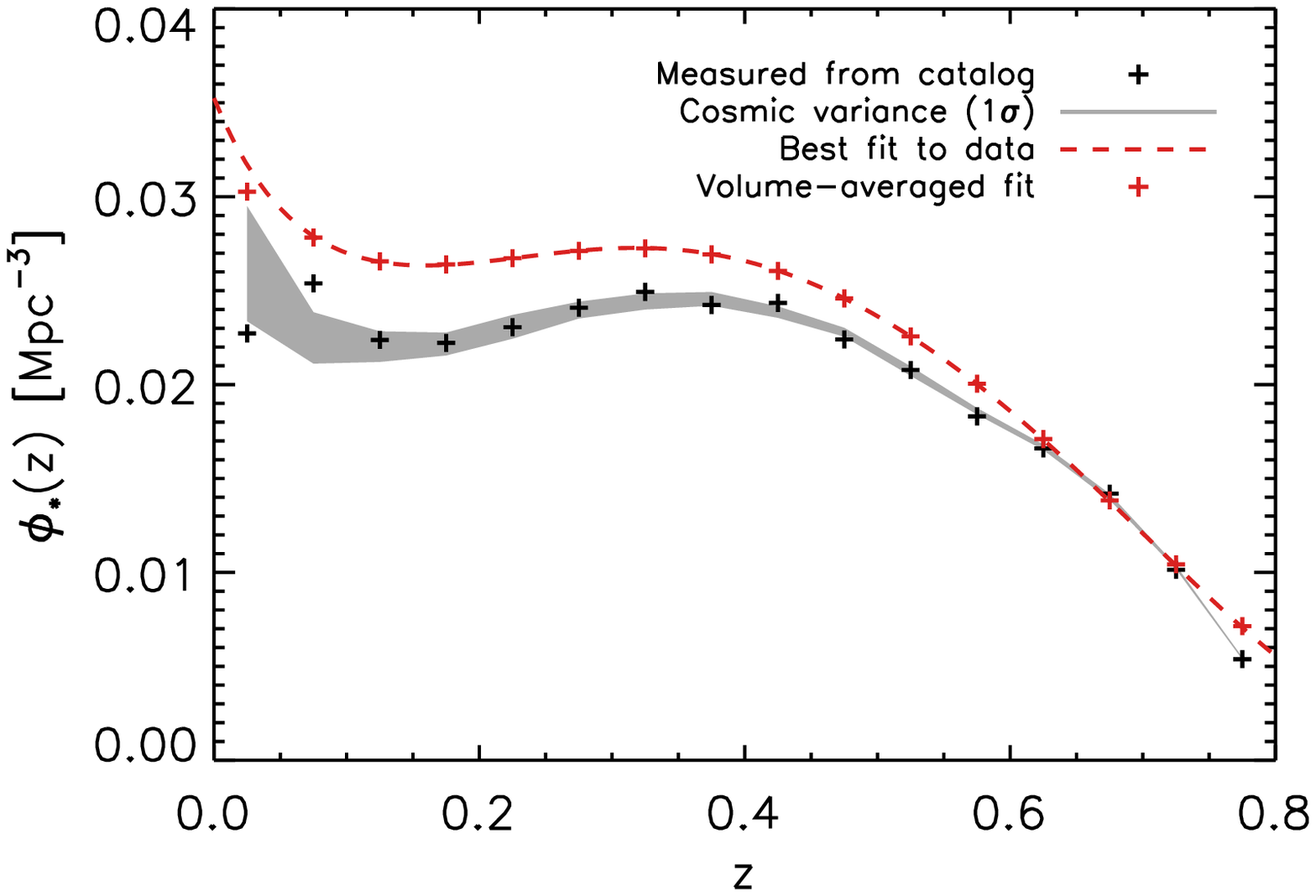}
\end{tabular}
\caption{The results for our fiducial model, minimized using equation \eqref{eq:chisq}. The model is constrained by two sets of data, one being the cross-correlation signal between photometric and spectroscopic galaxies (in bins of the apparent magnitude of the former and the redshift of the latter), the other being the total number of photometric galaxies in each bin of apparent magnitude. Lighter shaded bands show the effect of cosmic variance (see main text). \textit{Top left:} The number of galaxies in different bins of redshift as a function of absolute magnitude. Solid lines show the data, dashed lines show the outcome of the model. Note that the power-law part of the Schechter function is only probed by low-redshift galaxies. Overall the luminosity function of the data is reproduced very well. For the first redshift bin, where the deviation between the derived and true galaxy densities is largest, the vertical offset is in large part due to cosmic variance for the sky area we are using here (see main text). \textit{Top right:} The number of galaxies in different bins of redshift as a function of apparent magnitude. The total over all redshifts, shown by the black line, is one of the constraints of the model. \textit{Bottom left:} The number of galaxies in different bins of apparent magnitude as a function of redshift. Black lines show the total over all apparent magnitudes. \textit{Bottom right:} The normalization of the Schechter luminosity function as a function of redshift. The normalizations as inferred from the mock catalogue are shown as black crosses while what the best-fit model prefers is shown as a red dashed line. Red crosses show the result of volume-averaging the best fit over each redshift bin. Note that degeneracies exist between $\phi_*(z)$ and the other parameters of the Schechter function, which is why the number densities of the galaxies can be reproduced quite well for different sets of parameters.}
\label{fig:results_fid}
\end{center}
\end{figure*}

We note that the spectroscopic sample is (realistically) a biased subset of the galaxy population. As we show in Figure~\ref{fig:specvsphot}, the spectroscopic galaxies have a radically different redshift distribution and only probe the most luminous end of the total luminosity function. However, since both samples still trace the same large-scale distribution, and since the bias ratio of the two samples is a free parameter in the model, this is not an issue in our approach. Indeed, \citet{Scottez2016} recently showed that for the similar methodology of \citet{Menard2013}, accurate redshift distributions can be obtained for galaxies fainter than those of the spectroscopic sample. Our own results in the following section confirm this.

\subsection{Results}
\label{subsec:results}

\subsubsection{Fiducial model}
\label{subsubsec:fiducial}
Using only the clustering amplitude of the spectroscopic sources and the total distribution of photometric galaxies over apparent magnitude, our model is able to reproduce the input luminosity function of the mock catalogue to very high accuracy. The results for our fiducial model are shown in Figure~\ref{fig:results_fid}. Since error bars on the data shown or the model are not straightforwardly calculated, due to the many interrelated sources of uncertainty, we instead show just the $1\sigma$ variation due to cosmic variance on the data, as lightly shaded bands. This was calculated from $1000$ randomly placed surveys of the lightcone catalogue, each with the same sky area as our fiducial survey area.

\begin{figure*}
\begin{center}
\large \medskip\quad\quad\textbf{Direct maximum-likelihood fit}\par\medskip
\begin{tabular}{ccc}
\includegraphics[width=1.0\columnwidth, trim=22mm 8mm 8mm 8mm]{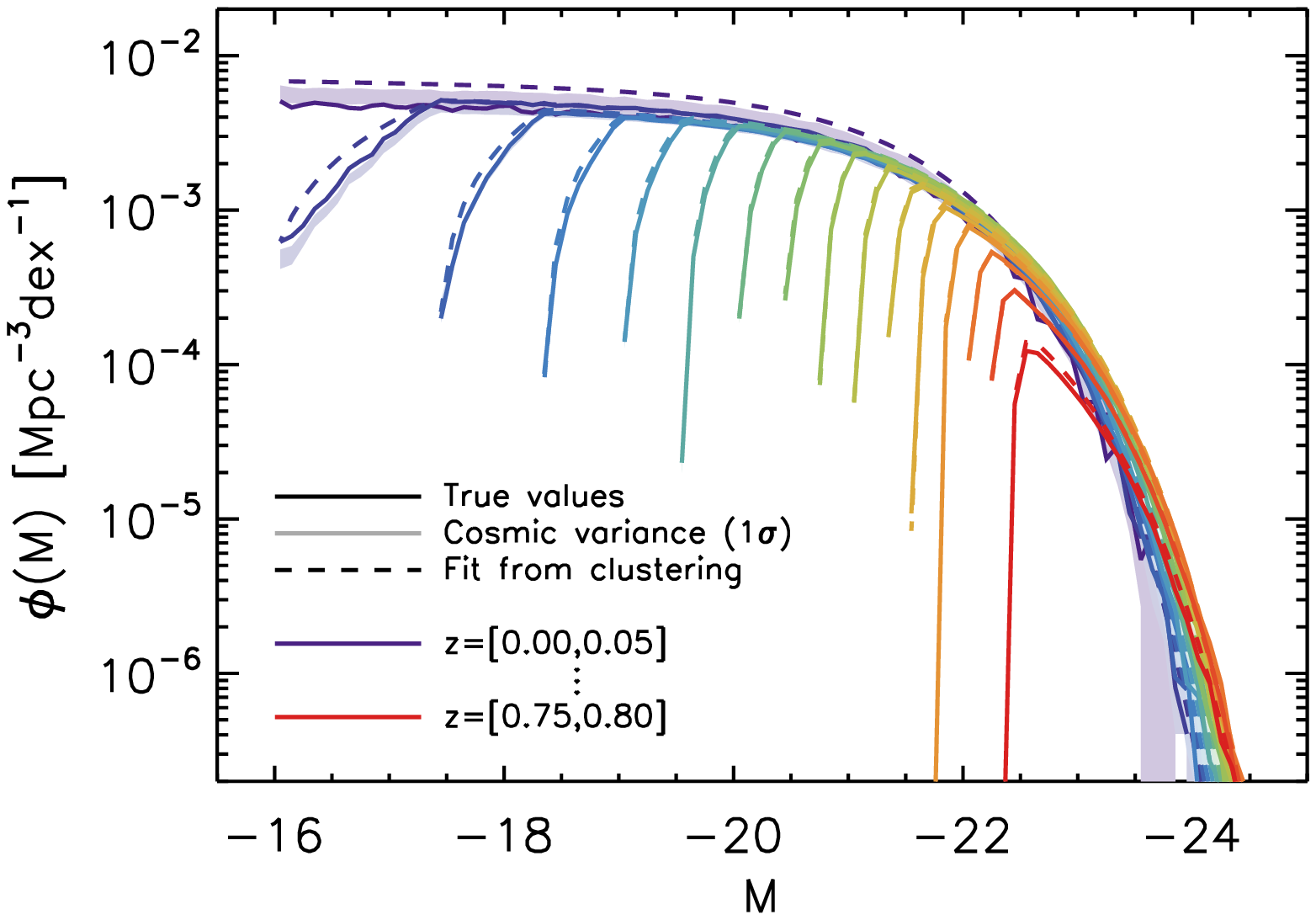} & &
\includegraphics[width=1.0\columnwidth, trim=22mm 8mm 8mm 8mm]{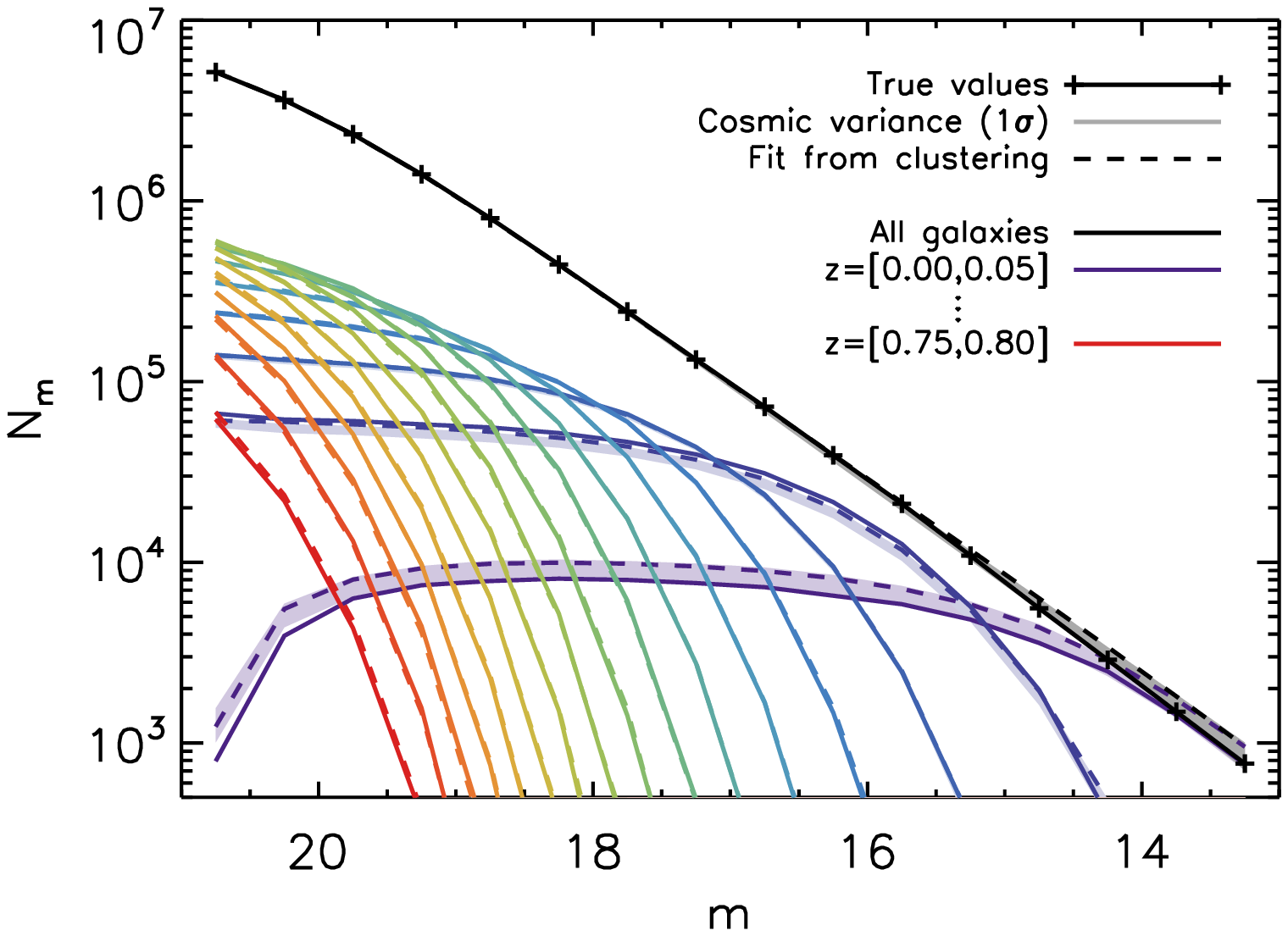}\\
\includegraphics[width=1.0\columnwidth, trim=22mm 8mm 8mm 8mm]{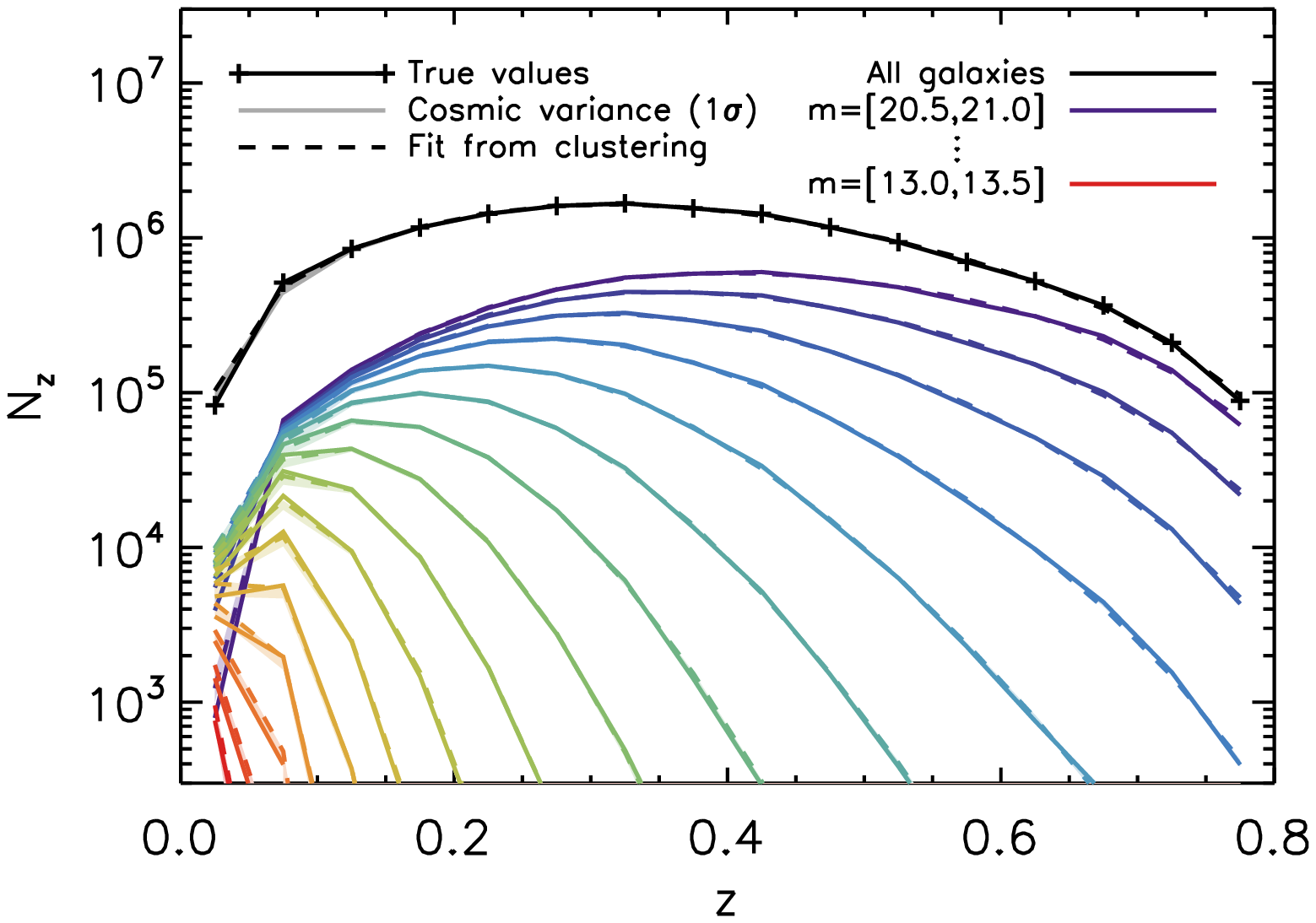} & &
\includegraphics[width=1.0\columnwidth, trim=22mm 8mm 8mm 8mm]{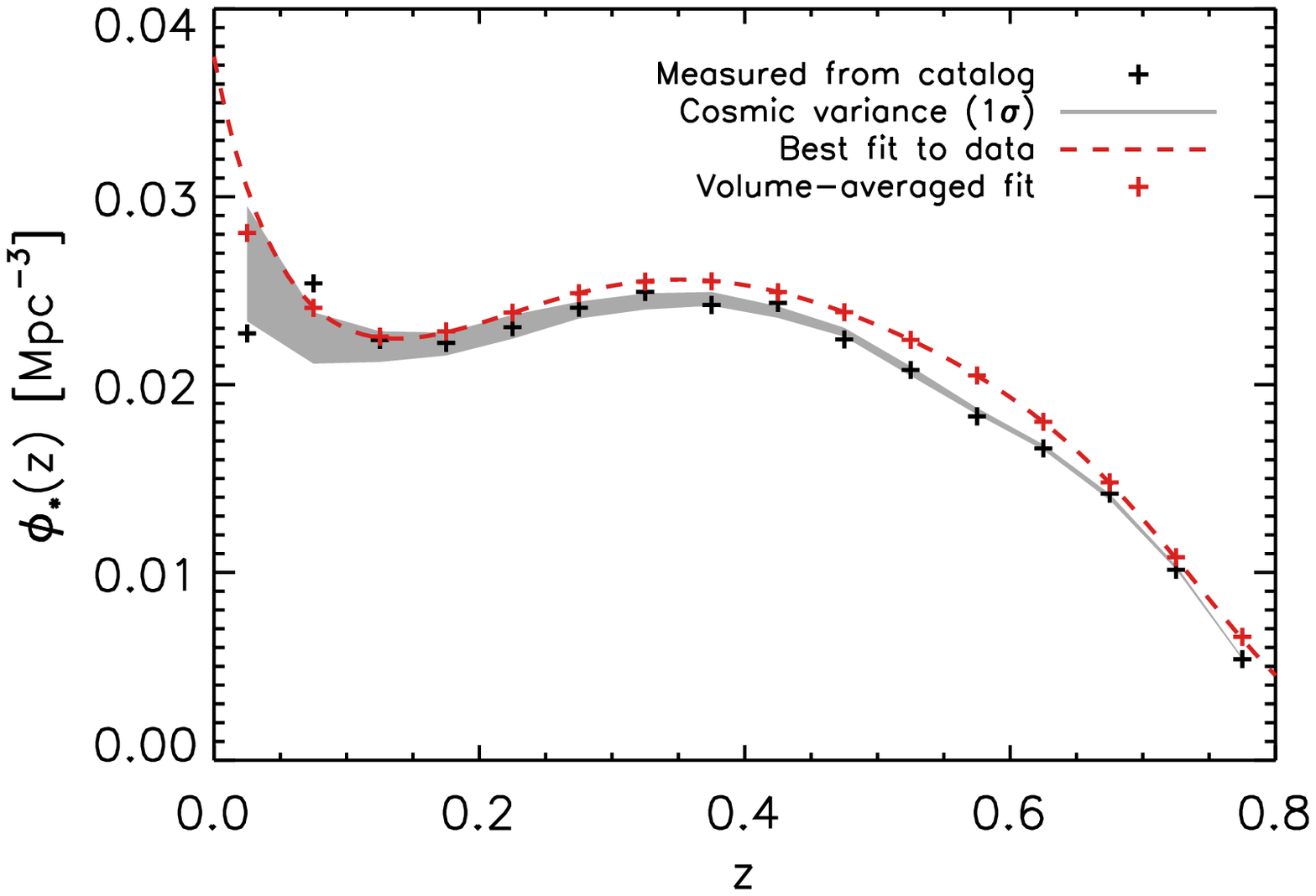}
\end{tabular}
\caption{As Figure~\ref{fig:results_fid}, but now the model has been replaced by a maximum-likelihood fit to the absolute magnitudes and redshifts of the mock galaxies, which are (realistically) inaccessible to the fiducial model. The luminosity functions derived here are extremely close to those of the fiducial model, showing that most of the already slight mismatch in Figure~\ref{fig:results_fid} is not due to the clustering signal or our clustering model, but to realization noise and the limitations of the parametrization of the luminosity function.}
\label{fig:results_dir}
\end{center}
\end{figure*}

\begin{figure*}
\begin{center}
\large \medskip\quad\quad\textbf{Known local slope}\par\medskip
\begin{tabular}{ccc}
\includegraphics[width=1.0\columnwidth, trim=22mm 8mm 8mm 8mm]{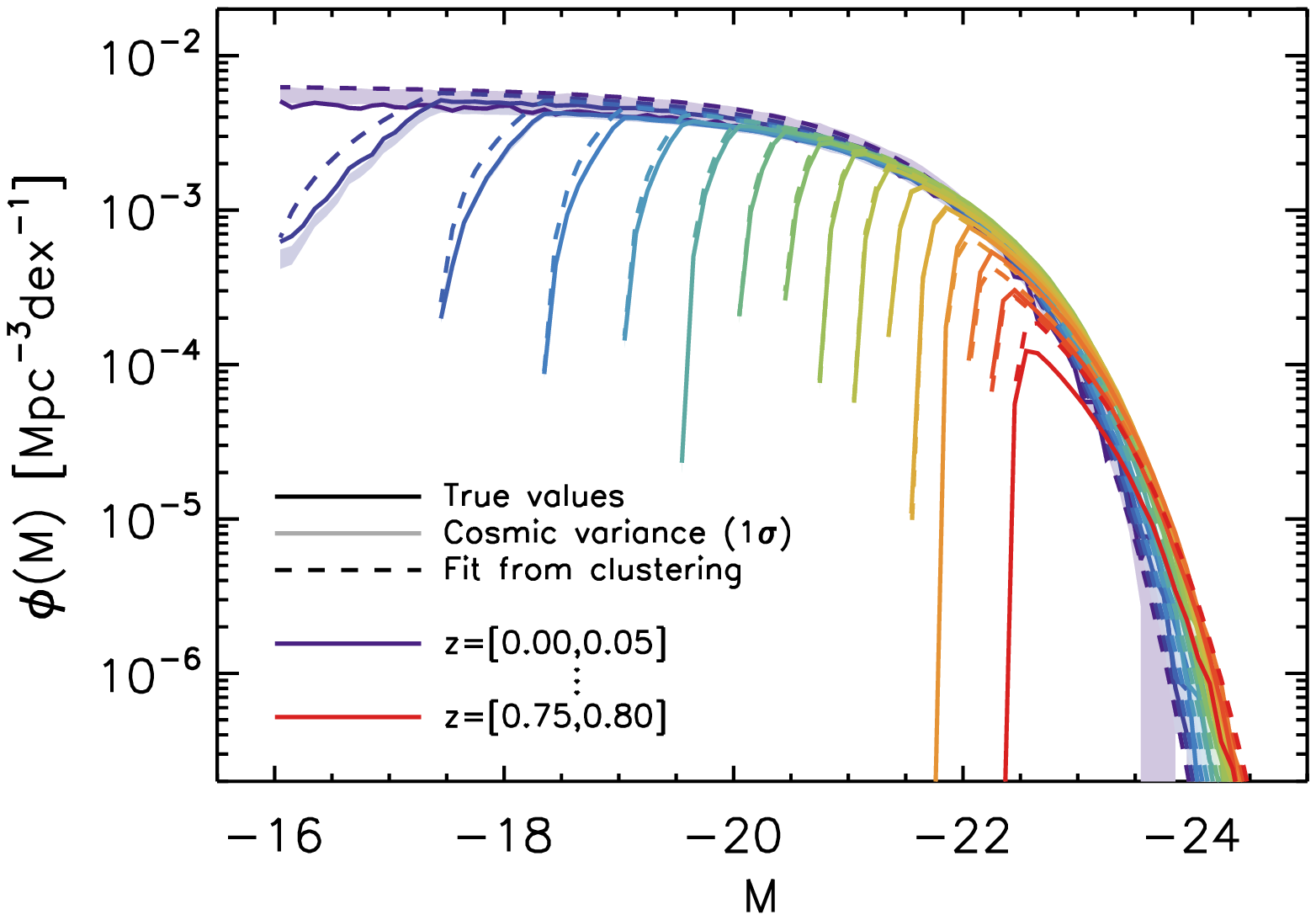} & &
\includegraphics[width=1.0\columnwidth, trim=22mm 8mm 8mm 8mm]{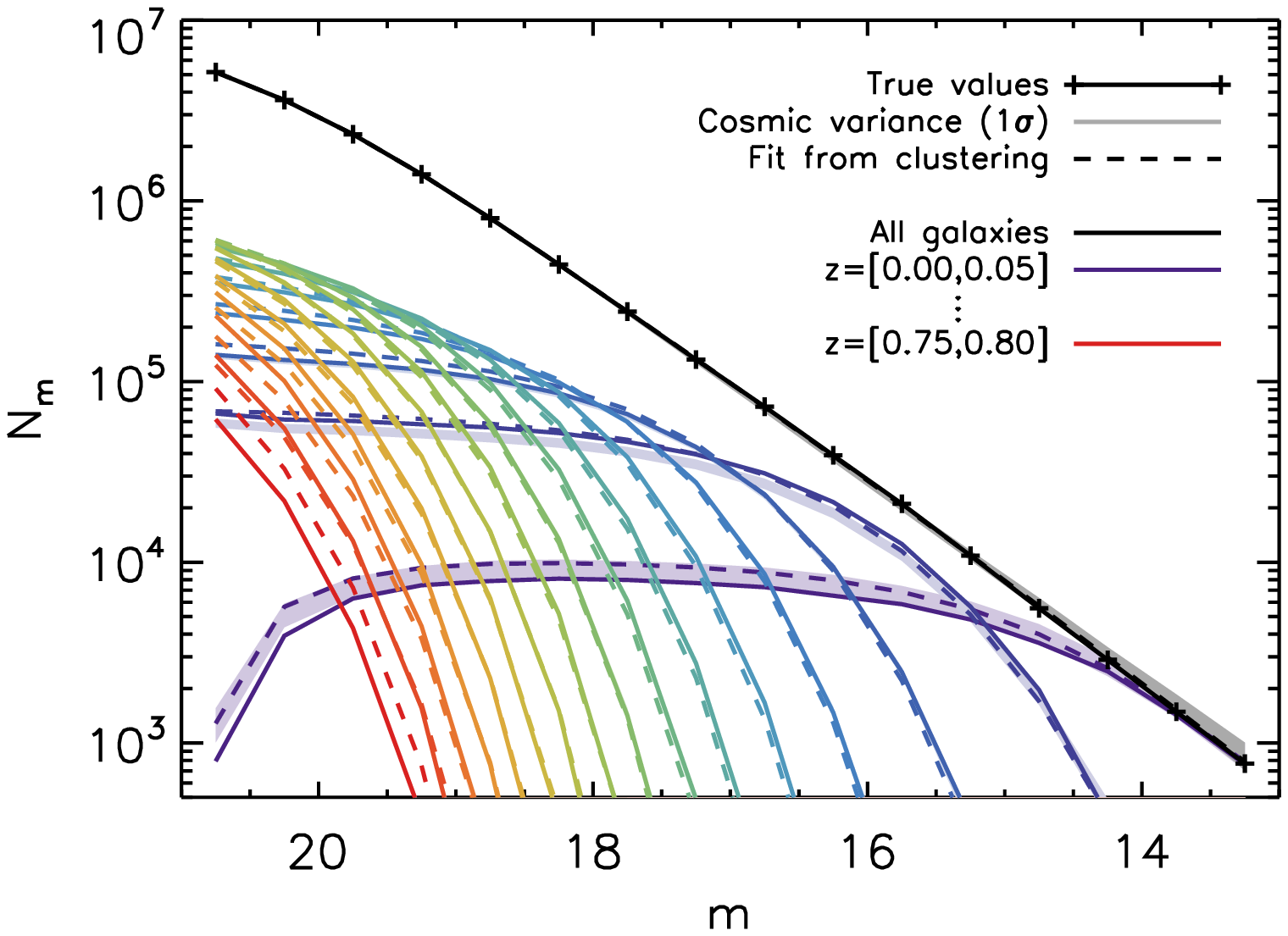}\\
\includegraphics[width=1.0\columnwidth, trim=22mm 8mm 8mm 8mm]{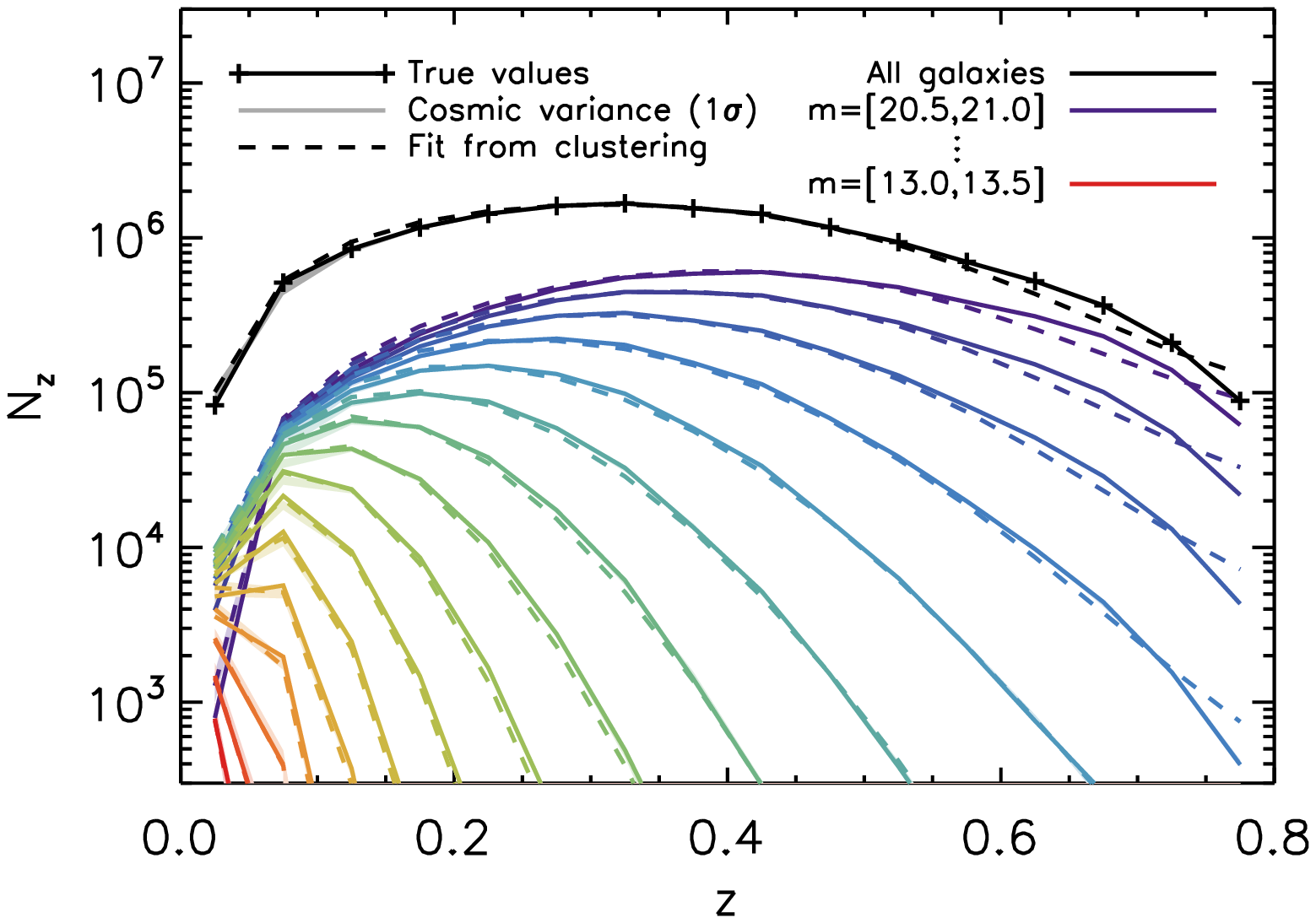} & &
\includegraphics[width=1.0\columnwidth, trim=22mm 8mm 8mm 8mm]{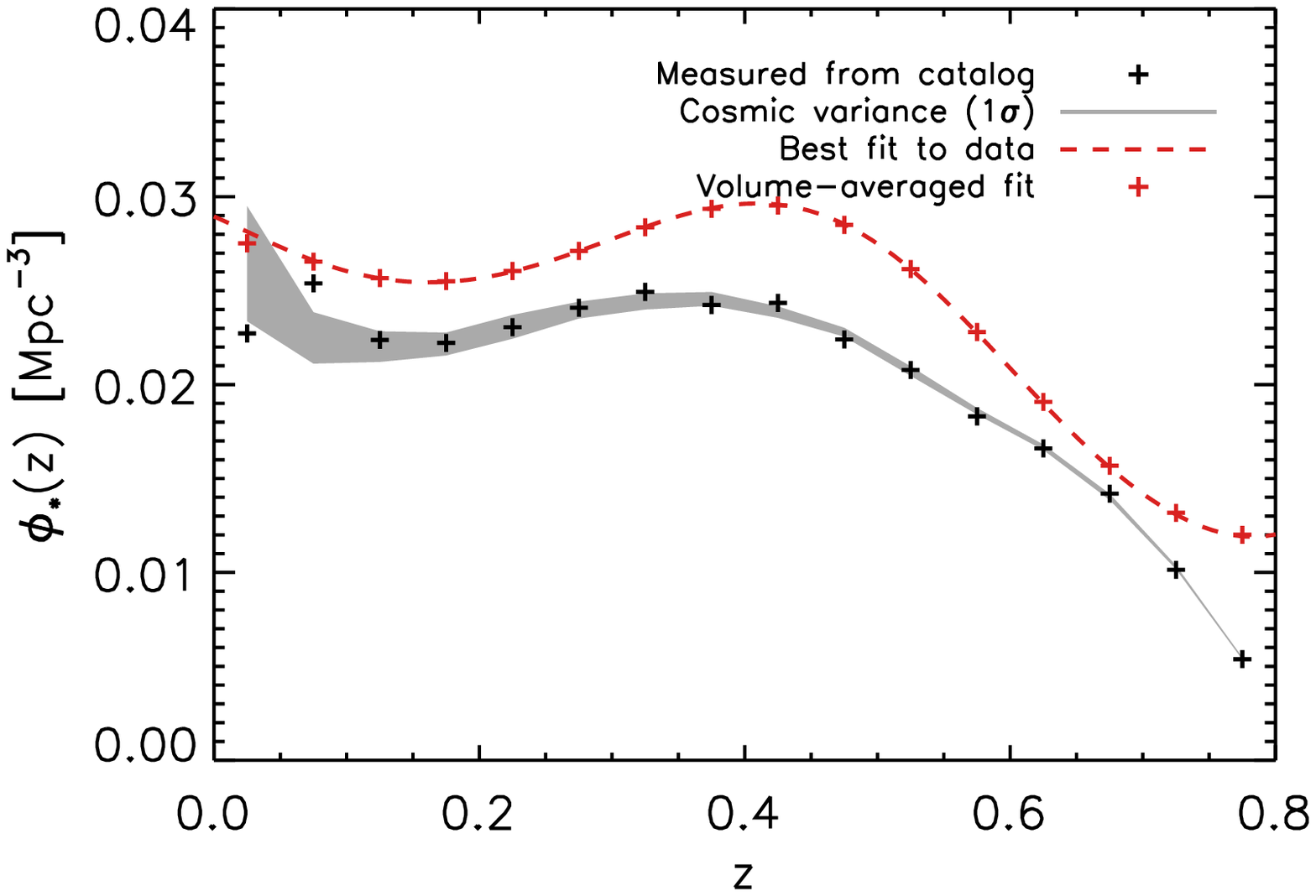}
\end{tabular}
\caption{As Figure~\ref{fig:results_fid}, but now $\alpha_0$, the power-law slope of the luminosity function at $z=0$, was assumed known and held fixed to the input value in the fitting. As expected, this marginally improves the fit at low redshift, but at the cost of some model freedom which is felt mainly at high redshift, where the fit worsens with respect to the fiducial model.}
\label{fig:results_a0}
\end{center}
\end{figure*}

In the top-left panel, we show the luminosity functions as a function of absolute magnitude in for each redshift bin. Solid lines show the luminosity function as measured directly from the mock catalogue with full redshift information, thereby including realisation noise (which plays a significant role in the first two redshift bins). At low redshift -- specifically in the first two redshift bins -- the model tends to overestimate the number of dim galaxies, although we note that the difference is in large part due to cosmic variance, as we will show. For the highest redshift bin, too, the model slightly overestimates the number of galaxies observed. Even with these caveats, in most regimes the luminosity function of the mock galaxies are very accurately reproduced by the best-fit galaxy distribution, including the bright and dim end dropoffs. The latter is due to the cut-off apparent magnitude shifting to brighter galaxies within each redshift bin, and therefore only captured when the model luminosity function and volume are integrated together (see \eqref{eq:Nlambdai_pre}).

We show the distribution over apparent magnitude for each redshift bin in the top-right panel. The total distribution is shown in black, and is used as a constraint in the model to break the clustering degeneracies (see equation~\eqref{eq:chisq}). The model again tends to overestimate the number of dim galaxies in the lowest redshift bin, where the cosmic variance is largest and the clustering signal has a relatively large uncertainty. Overall, though, the model does very well in reproducing the true distribution of galaxies in apparent magnitude, at any redshift.

The bottom-left panel of Figure~\ref{fig:results_fid} shows the redshift distributions in each apparent magnitude bin, as well as the total. Note that we are showing the absolute number of galaxies assigned to each redshift bin. The clustering model does an excellent job at reproducing these, even for the bright galaxies with relatively low number densities. As before, the fit is particularly accurate at intermediate redshifts (for all apparent magnitudes), where most of the galaxies in our sample reside and therefore where the uncertainty on the constraints is smallest.

\begin{figure*}
\begin{center}
\large \medskip\quad\quad\textbf{Limber approximation}\par\medskip
\begin{tabular}{ccc}
\includegraphics[width=1.0\columnwidth, trim=22mm 8mm 8mm 8mm]{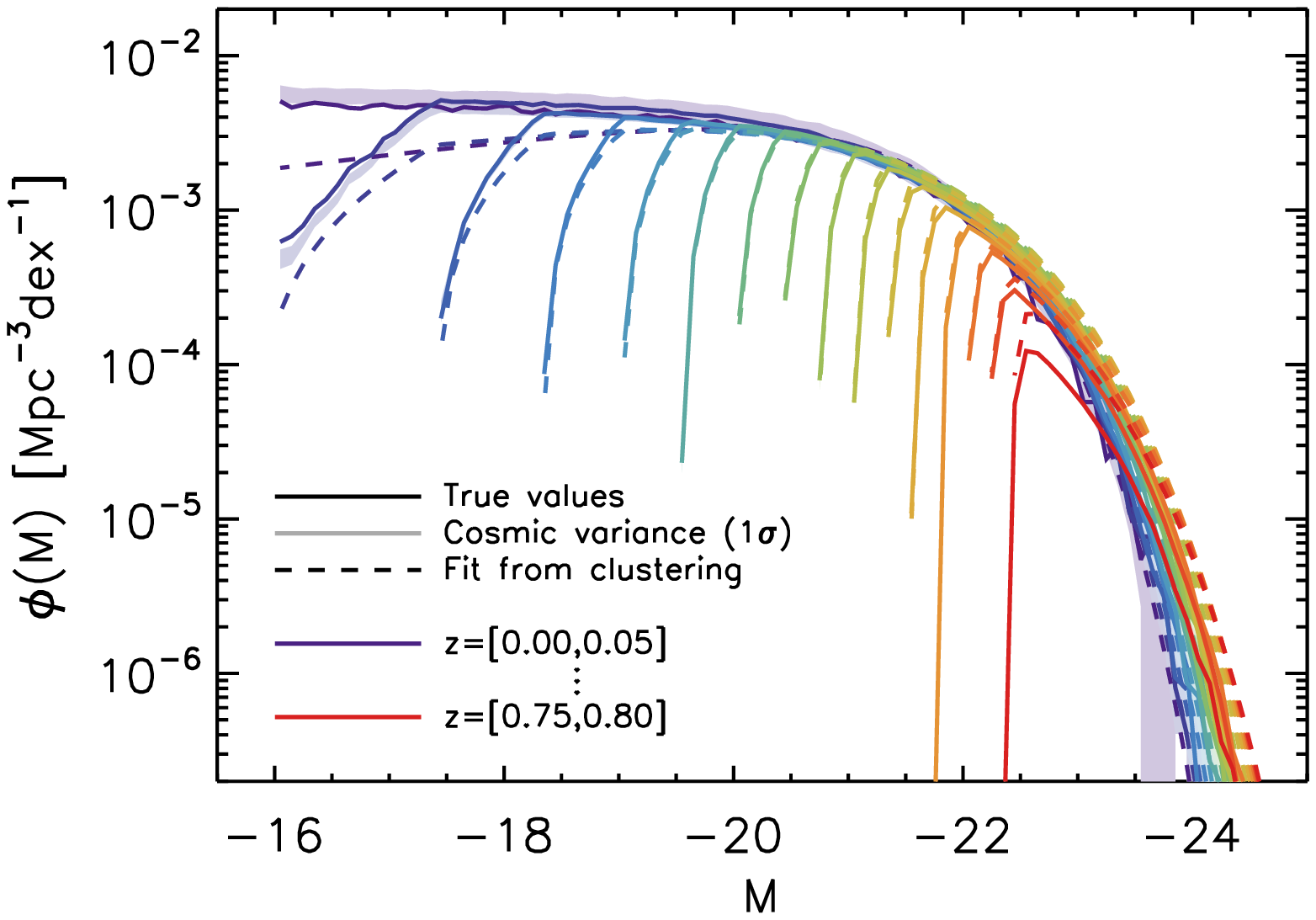} & &
\includegraphics[width=1.0\columnwidth, trim=22mm 8mm 8mm 8mm]{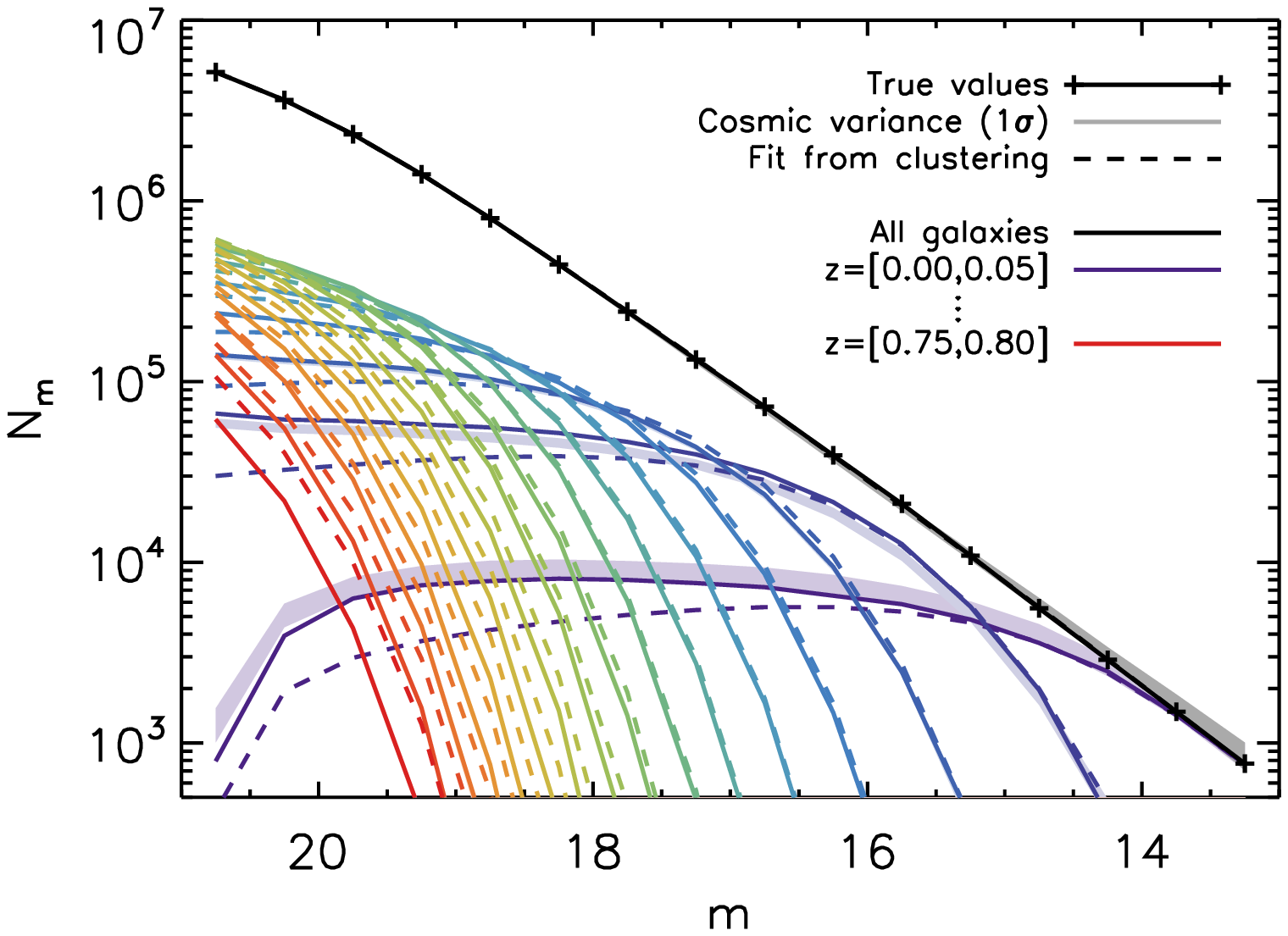}\\
\includegraphics[width=1.0\columnwidth, trim=22mm 8mm 8mm 8mm]{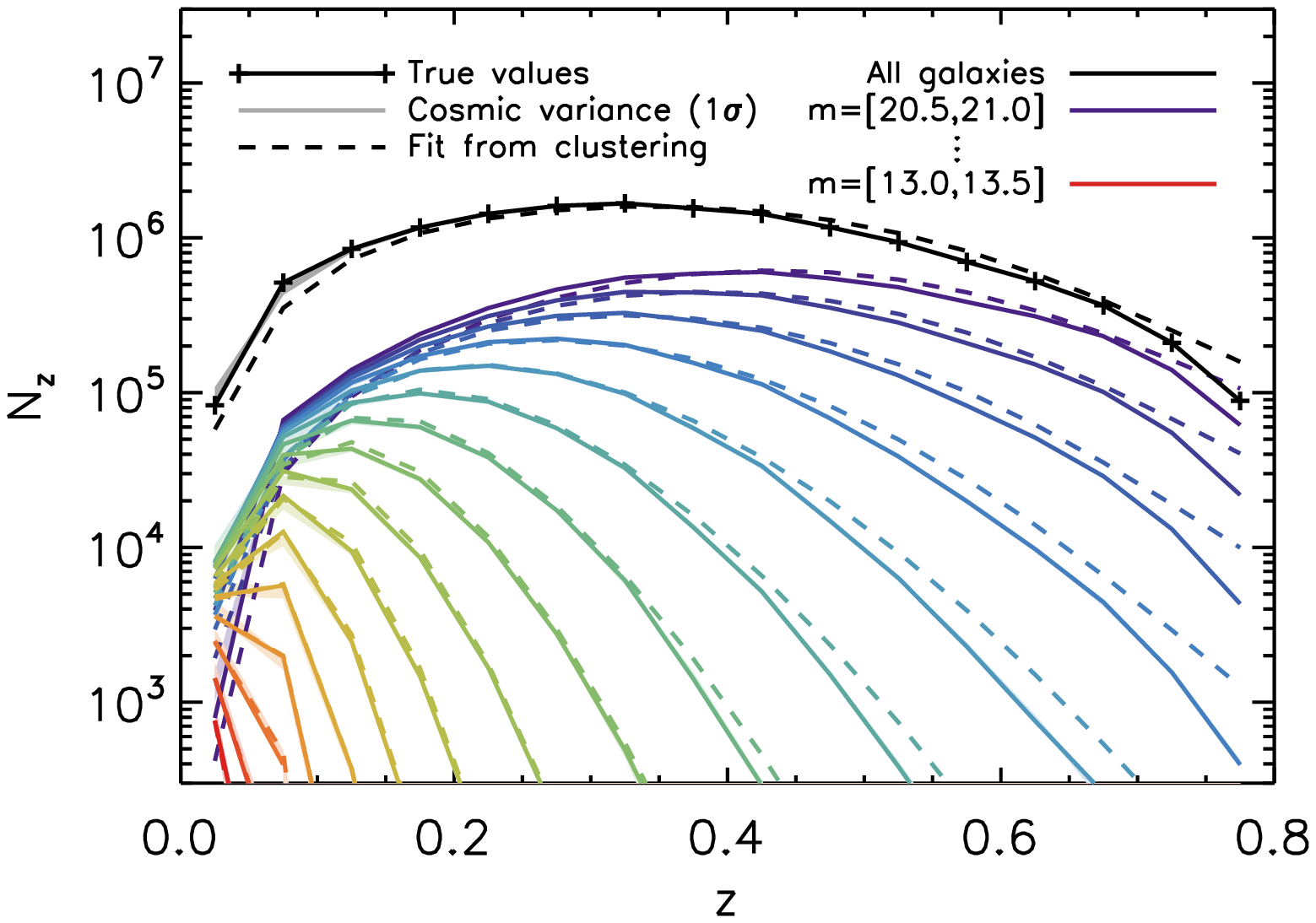} & &
\includegraphics[width=1.0\columnwidth, trim=22mm 8mm 8mm 8mm]{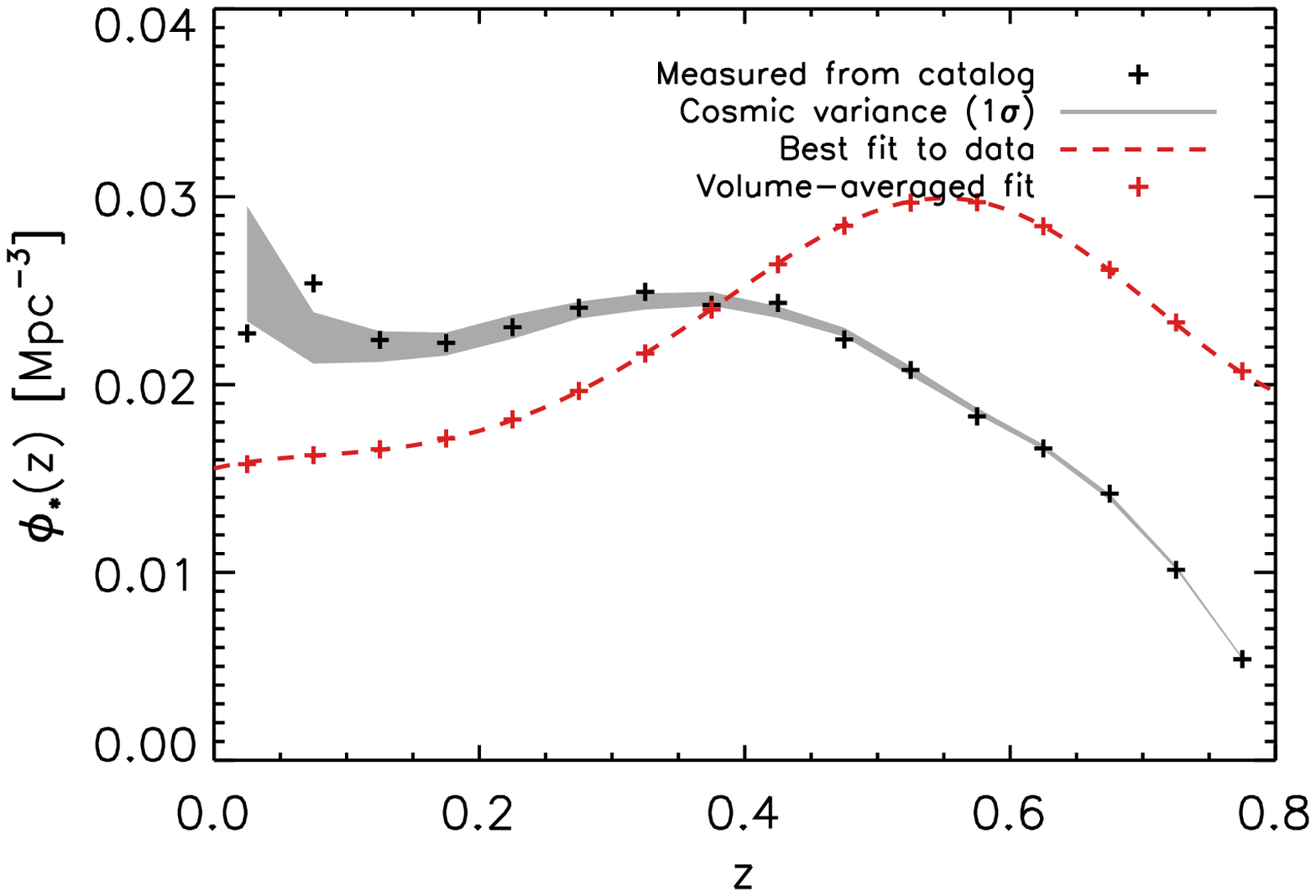}
\end{tabular}
\caption{As Figure~\ref{fig:results_fid}, but now with the Limber approximation taken. While the fit is still good at intermediate redshifts, both the luminosity function and $\dNdz$ are considerably less well reproduced at both low and high redshift. This is due to the small auto-correlation clustering amplitude and relatively high uncertainty at these redshifts, meaning contributions from cross-correlations between different redshift bins -- which are ignored in the Limber approximation -- are relatively more important.}
\label{fig:results_lim}
\end{center}
\end{figure*}

Finally, in the bottom-right panel, we show the normalization of the luminosity function as a function of redshift. Black crosses show the effective normalization of the mock galaxies in the survey area in each redshift bin. The dashed red line shows the fit (see equation~\eqref{eq:phistar}) that best reproduced the clustering data, with red crosses showing its volume-averaged values in each redshift bin to allow for a more direct comparison to the input data. The fit captures the shape of the input data, even if it tends to overestimate the normalization. However, due to the degeneracies between different Schechter parameters\footnote{One easily seen example of such a degeneracy is between the high-redshift normalization and the slope parameters of the Schechter function. At high redshifts, galaxies above the knee ($M>M_*$) are not or barely probed as they are too dim to observe, and so in this regime the slope parameters only serve to normalize the profile.}, a mismatch in the value of the normalization parameter does not necessarily mean that the luminosity function itself is not accurately reproduced, as the other panels show.

If we compare the normalization measured for our catalogue to the shaded band showing the $1\sigma$ range of cosmic variance, we see that our survey area contains significantly less galaxies than average in the first redshift bin, and significantly more than average in the second redshift bin. This uncommon feature is the main reason why our model has trouble matching the measured number densities in these redshift bins. Also note the sharp downturn to lower number densities observed for the very highest redshift bin, which is not fully captured by our fit, causing the model to overestimate the number of galaxies in that bin.

\subsubsection{Direct maximum-likelihood fit}
\label{subsubsec:direct}
To show that the mismatch at low redshift is indeed not due to the clustering signal or shortcomings of our clustering model, we show in Figure~\ref{fig:results_dir} the results of performing a maximum-likelihood fit directly to the absolute magnitudes and redshifts of the galaxies in the survey catalogue, both of which our fiducial model is agnostic about. We do not bin the data here, instead using the individual $M$ and $z$ of each galaxy as input to the maximum-likelihood function (see Appendix~\ref{app:directlike}). Even in this case, the number of galaxies at low redshift is overpredicted, due to realization noise (which includes cosmic variance). Comparing Figures \ref{fig:results_fid} and \ref{fig:results_dir}, we see that the result of our fiducial model is extremely close to the maximum-likelihood luminosity function, showing the power of using the cross-correlation signal even without any prior redshift information. Additionally, this shows that that the cumulative impact of binning, uncertainties in the clustering data, and perhaps most significantly our assumptions regarding the clustering bias, is small.

Adding more parameters to the luminosity function, by for example including a second Schechter function or higher-order terms in the normalization, would allow us to compensate for the realization noise and possibly yield a better match to the data. However, doing so would also introduce additional degeneracies.

\subsubsection{Fixed slope at low redshift}
\label{subsubsec:fixeda0}
Our fiducial model has no prior information on the parameters of the luminosity function. However, it is not unreasonable to assume that the power-law slope of the luminosity function at redshift zero, $\alpha_0$, is well-constrained. To see how much the model outcome is influenced by the uncertainty at low redshift, we therefore also ran our model with $\alpha_0$ fixed to the input value. The results of this test are shown in Figure~\ref{fig:results_a0}. As expected, the panels show a marginal improvement at low redshift in comparison to the results for our fiducial model, but our results at high redshift are slightly worse than before. This is again because of the unusually large realization noise at low redshift: as one parameter is held fixed, the model loses some freedom to compensate for this, which in this case leads to a mismatch at high redshift.

\subsubsection{Limber approximation}
\label{subsubsec:limber}
Finally, we have also tested the consequences of assuming the often-used Limber approximation, by setting the clustering signal (and its covariance) to zero for the cross-correlations of spectroscopic sources in different redshift, the results of which are shown in Figure~\ref{fig:results_lim}. In this case, the model performs less well in regimes where the cross-correlations between different redshift bins contribute significantly -- that is, at both the low and high redshift ends, and for the brightest galaxies, which have relatively low number densities. At the lowest redshift, depending on the choice of $\theta_\mathrm{max}$ (see \S\ref{subsec:crosscorr}) the typical distance between galaxies may be larger than the distances probed by the clustering signal, and so no or barely any clustering is observed. Without the information contained in the cross-correlation signal between these and higher-redshift bins, the model therefore prefers to place as few galaxies as possible at low redshift. At high redshift, depending on the choice of $\theta_\mathrm{min}$ the scales probed may be larger than the scales on which those galaxies cluster strongly, and so a weak signal with a relatively large uncertainty is observed. Increasing $\theta_\mathrm{max}$/decreasing $\theta_\mathrm{min}$ gives better results at low/high redshifts but increases the uncertainty at higher/lower redshifts. It is therefore best to not take the Limber approximation but make use of all available information. If the Limber approximation has to be taken, it is better to calculate the clustering at a fixed physical scale instead of a fixed angular scale \citep[e.g.][]{Schulz2010}.\newline

\noindent
For completeness, we present the best-fit Schechter parameters corresponding to all figures in this section in Table~\ref{tab:params}. Note that the reproduced luminosity functions can be quite accurate even when the parameters are not, because of the degeneracies of some of these parameters with the normalization.

\begin{table}
\centering
\begin{tabular}{l c c c c}
\hline
Run & $\alpha_0$ & $M_{*0}$ & $\alpha_\mathrm{e}$ & $M_\mathrm{*e}$ \\ [0.5ex]
\hline
\textbf{Input} & \textbf{-1.01} & \textbf{-21.5} & \textbf{-0.15} & \textbf{-0.8} \\ [1ex]
Fiducial & -1.050 & -21.429 & -0.155 & -0.927 \\ [1ex]
Direct & -1.019 & -21.520 & -0.178 & -0.783 \\ [1ex]
Fixed $\alpha_0$ & (-1.01) & -21.415 & -0.337 & -1.061 \\ [1ex]
Limber & -0.747 & -21.349 & -0.854 & -1.344 \\ [1ex]
\hline
\end{tabular}
\caption{The best-fit luminosity function parameters derived from the clustering data for each of our model runs. Parentheses indicate that the parameter was held fixed to this value. In the run labelled ``Direct'' no clustering information was used, and instead a maximum-likelihood fit to the galaxies absolute magnitudes and redshifts was performed. Note that the luminosity function may be highly accurately reproduced even for parameters other than the input parameters, due to degeneracy with the normalization and realization noise (including cosmic variance).}
\label{tab:params}
\end{table}

\section{Discussion}
\label{sec:discussion}
The methods presented in this paper extend previous work by not only deriving the redshift distribution of photometric sources through clustering, but also their luminosity function through cosmic time. By testing this method on a mock galaxy survey, we have demonstrated that an input galaxy distribution over redshift and luminosity can be very accurately recovered in this way for large surveys, even when these are relatively shallow. The redshift distributions derived in this way are not biased by having the spectroscopic sources be selected differently from the photometric sources. As we have shown, the method returns accurate distributions and luminosity functions even if the only galaxies with spectra are the brightest members of the sample and their number densities have a vastly different redshift evolution, so long as they are in the same area of sky. Additionally, our results are not degenerate with the unknown redshift-dependent galaxy bias, $b(z)$.

Our goal has been to introduce a technique for measuring the luminosity function from the co-spatial combination of a deep imaging survey and a sparse spectroscopic survey and to illustrate its potential. The performance of our simple algorithm on mock data is sufficiently encouraging that further development appears warranted. In particular, application to real data would need to consider the possible effects of lensing magnification and incorporate K-corrections in the conversion between apparent and absolute magnitudes (see Appendix~\ref{app:ignored} for more on this). Additionally, in this paper we have taken the following assumptions, which should be kept in mind and modified where necessary:

\begin{itemize}
\item First of all, we have assumed that the form of the luminosity function is known (in our case, a single Schechter function), which in real surveys may not be the case. However, one generally finds that a sum of Schechter functions is a good fit to real data \citep[e.g.][]{Peng2010}. Additionally, the form of the luminosity function that one assumes in this formalism can be very versatile, and is allowed to contain many parameters to be constrained at once. We therefore do not anticipate this to be an issue in the application of the model.

\item Second, we have assumed a simple luminosity bias relation (equation~\eqref{eq:bL}) with a known parameter $L'$. We have also assumed that the redshift evolution of the remaining bias terms cancel out. However, we have imposed neither bias relation on the mock data, and our results imply these assumptions were sufficiently valid. There is no reason to assume, therefore, that the same would not apply to real data -- except perhaps if the clustering bias in the real data had some residual dependence on redshift and/or magnitude that the mock data does not. Any potential scale-dependence of the clustering bias (insofar not already implicitly included in the mock data) is not expected to be important, as the bias in our model is an effective one, averaged over a large range in scales. Finally, while the value of $L'$ was fit to a subset of the data prior to running the model, it could in principle be a free parameter constrained by the model.

\item Third, as we mentioned in \S\ref{subsec:modelfit}, it is difficult to define an objective value for the relative weight $R$ of the two terms in our model's $\chi^2$, in equation~\eqref{eq:chisq}. Fortunately, the outcome of the model turns out not to be very sensitive to its value.
\end{itemize}

We plan to test our method on a large catalogue of observed galaxies in a follow-up publication.

\section*{Acknowledgements}
The authors thank Joanne Cohn for helpful discussions, and Yu Feng for the use of his clustering code. We also thank the anonymous referee for their useful comments, which have led to the improvement of this manuscript. This work was supported in part by the Theoretical Astrophysics Center at UCB. The Millennium Simulation databases used in this paper and the web application providing online access to them were constructed as part of the activities of the German Astrophysical Virtual Observatory.
\bibliographystyle{mn2e}
\setlength{\bibhang}{2.0em}
\setlength{\labelwidth}{0.0em}
\bibliography{zlfbib}

\appendix
\section{Joint covariance matrix}
\label{app:covar}
There are three sources of uncertainty when fitting our model to the data: uncertainties in the integrated cross-correlation function of photometric and spectroscopic galaxies, $\wps{}_{,\lambda i} \equiv \wps(m_\lambda,z_i)$, in the integrated cross-correlation function of spectroscopic galaxies in different redshift bins, $\wss{}_{,ij} \equiv \wss(z_i,z_j)$, and finally in the number of galaxies in the volume at some apparent magnitude and redshift, $\Npm{}_{,\lambda i} \equiv \Npm(m_\lambda,z_i)$. For the first two, we use $20,\!000$ bootstrap resamplings to calculate full covariance matrices, while the latter is modelled as a Poisson variable with a mean given by the volume-weighted integral over the luminosity function over bins $m_\lambda$ and $z_i$. Here we derive the total covariance matrix, which incorporates the uncertainties from all three sources.

To find the best-fit model, we aim to minimize $\chi^2$ as given by equation~\eqref{eq:chisq}, where $C$ is the joint covariance matrix. As such, $C$ is a $(n_\mathrm{m} n_\mathrm{z})\times(n_\mathrm{m} n_\mathrm{z})$ matrix with element $((\lambda i),(\mu j))$ given by:
\begin{eqnarray}
\label{eq:Cij1}
C_{(\lambda i)(\mu j)} \!\!\!\!\!&=&\!\!\!\! \sigma\left(\wps{}_{,\lambda i}-\wpsm{}_{,\lambda i} \,;\, \wps{}_{,\mu j}-\wpsm{}_{,\mu j}\right)\\
\nonumber
\!\!\!\!\!&=&\!\!\!\! \sigma\!\!\left(\wps{}_{,\lambda i}-\sum_k X_{ik}\,\fN'{}_{,\lambda k} \,;\, \wps{}_{,\mu j}-\sum_l X_{jl}\,\fN'{}_{,\mu l}\right)\!,
\end{eqnarray}
where $\sigma(A \,;B)$ denotes the covariance between $A$ and $B$. Note that $C$ is symmetric. As before, $X_{ij}=\wss{}_{,ij}$ and $\fN'{}_{,\lambda i}=K\bL{}_{,\lambda i} \Npm{}_{,\lambda i}/\Npm{}_{,\lambda}$, where $K$ is a constant, $\bL{}_{,\lambda i}$ is the part of the bias that scales with the luminosity of a galaxy of apparent magnitude $m_\lambda$ at redshift $z_i$ (see equation \eqref{eq:bL}) and $\Npm{}_{,\lambda}=\sum_i \Npm{}_{,\lambda i}$ is the total number of galaxies observed in apparent magnitude bin $m_\lambda$. While $\Npm{}_{,\lambda}$ and $\bL{}_{,\lambda i}$ are known \textsl{a priori}, $K$ is a parameter of the model. Expanding equation~\eqref{eq:Cij1}, we find:
\begin{eqnarray}
\label{eq:Cij2}
\nonumber
C_{(\lambda i)(\mu j)} \!\!\!\!\!&=&\!\!\!\! \sigma(\wps{}_{,\lambda i} \,;\, \wps{}_{,\mu j}) - \\
\nonumber
\!\!\!\!\!& &\!\!\!\!\sum_l \bL{}_{,\mu l}\,\sigma\!\!\left(\wps{}_{,\lambda i} \,;\, \frac{\Npm{}_{,\mu l}}{\Npm{}_{,\mu}}\,\wss{}_{,jl}\right) - \\
\!\!\!\!\!& &\!\!\!\!K\sum_k \bL{}_{,\lambda k}\,\sigma\!\!\left(\wps{}_{,\mu j} \,;\, \frac{\Npm{}_{,\lambda k}}{\Npm{}_{,\lambda}}\,\wss{}_{,ik}\right) + \\
\nonumber
\!\!\!\!\!& &\!\!\!\!K^2\sum_{k,l} \bL{}_{,\lambda k} \bL{}_{,\mu l}\,\sigma\!\!\left(\frac{\Npm{}_{,\lambda k}}{\Npm{}_{,\lambda}}\,\wss{}_{,ik} \,;\, \frac{\Npm{}_{,\mu l}}{\Npm{}_{,\mu}}\,\wss{}_{,jl}\right)\!.
\end{eqnarray}
It is clear that $\wss{}_{,ij}$ and $\Npm{}_{,\lambda k}$ should be uncorrelated, and we assume the same for $\Npm{}_{,\lambda k}$ and $\wps{}_{,\mu i}$. With this in mind, we can write:
\begin{equation}
\label{eq:Cij3}
\sigma\!\!\left(\wps{}_{,\lambda i} \,;\, \frac{\Npm{}_{,\mu l}}{\Npm{}_{,\mu}}\,\wss{}_{,jl}\right) = \frac{\Npm{}_{,\mu l}}{\Npm{}_{,\mu}}\,\sigma(\wps{}_{,\lambda i} \,;\, \wss{}_{,jl}),
\end{equation}
and:
\begin{eqnarray}
\nonumber
\sigma\!\!\left(\frac{\Npm{}_{,\lambda k}}{\Npm{}_{,\lambda}}\right.\!\!\!\!\!\!&\,&\!\!\!\!\!\!\!\left.\wss{}_{,ik} \,;\, \frac{\Npm{}_{,\mu l}}{\Npm{}_{,\lambda}}\,\wss{}_{,jl}\right) = \\
\label{eq:Cij4}
\!\!\!\!\!\!& &\!\!\!\!\!\! \frac{\Npm{}_{,\lambda k}\Npm{}_{,\mu l}}{\Npm{}_{,\lambda}\Npm{}_{,\mu}}\,\sigma(\wss{}_{,ik} \,;\, \wss{}_{,jl}) + \\
\nonumber
\!\!\!\!\!\!& &\!\!\!\!\!\! \left[\wss{}_{,ik}\wss{}_{,jl} + \sigma(\wss{}_{,ik} \,;\, \wss{}_{,jl})\right]\sigma\!\!\left(\frac{\Npm{}_{,\lambda k}}{\Npm{}_{,\lambda}} \,;\, \frac{\Npm{}_{,\mu l}}{\Npm{}_{,\mu}}\right)\!.
\end{eqnarray}
All remaining covariances involving the clustering terms are calculated directly through bootstrapping. This leaves only the last covariance in equation~\eqref{eq:Cij4}. The $\Npm{}_{,\lambda i}$ are mutually independent Poisson variables, but are not independent of $\Npm{}_{,\mu}$ when $\mu=\lambda$. So:
\begin{eqnarray}
\nonumber
C_{(\lambda i)(\mu j)} \!\!\!\!\!&=&\!\!\!\! \sigma(\wps{}_{,\lambda i} \,;\, \wps{}_{,\mu j}) - \\
\nonumber
\!\!\!\!\!& &\!\!\!\!\sum_k \fN'{}_{,\lambda k}\,\sigma(\wps{}_{,\mu j} \,;\, \wss{}_{,ik}) - \\
\label{eq:Cijfinal}
\!\!\!\!\!& &\!\!\!\!\sum_l \fN'{}_{,\mu l}\,\sigma(\wps{}_{,\lambda i} \,;\, \wss{}_{,jl}) + \\
\nonumber
\!\!\!\!\!& &\!\!\!\!\sum_{k,l} \fN'{}_{,\lambda k}\fN'{}_{,\mu l}\,\sigma(\wss{}_{,ik} \,;\, \wss{}_{,jl}) + \\
\nonumber
\!\!\!\!\!& &\!\!\!\!\delta_{\lambda\mu}\sum_{k,l} K^2\bL{}_{,\lambda k}\bL{}_{,\lambda l}\left[\wss{}_{,ik}\wss{}_{,jk}\, + \right.\\
\nonumber
\!\!\!\!\!& &\!\!\!\!\left.\sigma(\wss{}_{,ik} \,;\, \wss{}_{,jk})\right]\sigma\!\!\left(\frac{\Npm{}_{,\lambda k}}{\Npm{}_{,\lambda}} \,;\, \frac{\Npm{}_{,\lambda l}}{\Npm{}_{,\lambda}}\right)\!.
\end{eqnarray}
Since $\Npm{}_{,\lambda}$ is a sum of independent Poisson variables, and therefore a Poisson distributed variable itself, we need to know the covariance between ratios of dependent Poisson variables in the domain $[0,1]$. Analytical expressions for this (co)variance and its derivatives can be derived, and the former are given below for completeness. Here $\gamma$ is Euler's constant, and $\mathrm{Ei}$ is the exponential integral function.

If $k=l$:
\begin{eqnarray}
\sigma^2\left(\frac{\Npm{}_{,\lambda k}}{\Npm{}_{,\lambda}}\right) \!\!\!\!\!&=&\!\!\!\!\! \frac{\Npm{}_{,\lambda k}}{\Npm{}_{,\lambda}^2}e^{-\Npm{}_{,\lambda}}\left\{\Npm{}_{,\lambda k}\left(1-e^{-\Npm{}_{,\lambda}}\right) + \right.\\
\nonumber
\!\!\!\!\!& &\!\!\!\!\! \left.\left(\Npm{}_{,\lambda}-\Npm{}_{,\lambda k}\right)\left(\mathrm{Ei}\left[\Npm{}_{,\lambda}\right]-\gamma-\ln\left[\Npm{}_{,\lambda}\right]\right)\right\}.
\end{eqnarray}

In all other cases:
\begin{eqnarray}
\nonumber
\sigma\!\!\left(\frac{\Npm{}_{,\lambda k}}{\Npm{}_{,\lambda}} \,;\, \frac{\Npm{}_{,\lambda l}}{\Npm{}_{,\lambda}}\right) \!\!\!\!\!&=&\!\!\!\!\! \frac{\Npm{}_{,\lambda k}\Npm{}_{,\lambda l}}{\Npm{}_{,\lambda}^2}e^{-\Npm{}_{,\lambda}}\left(1-e^{-\Npm{}_{,\lambda}} - \right.\\
\!\!\!\!\!& &\!\!\!\!\! \left.\vphantom{e^{-\Npm{}_{,\lambda}}}\mathrm{Ei}\left[\Npm{}_{,\lambda}\right]+\gamma+\ln\left[\Npm{}_{,\lambda}\right]\right)\!.
\end{eqnarray}

\section{Direct likelihood function}
\label{app:directlike}
In order to test our clustering-based approach to finding the galaxy redshift distribution and luminosity function, as well as test the cumulative impact of binning, the uncertainties of the clustering data and our model choices regarding the clustering data (e.g.\ the bias model), in \S\ref{subsubsec:direct} we considered the luminosity function one would obtain when doing a direct maximum-likelihood fit to the individual absolute magnitudes and redshifts of the galaxies. The likelihood function we maximised is constructed as follows.

We assume that the set of observed galaxies is a Poisson realization, with Poisson means determined by an underlying luminosity function and cosmology (see equation~\eqref{eq:Nlambdai}). Let us now consider these Poisson means in bins in $z$ and $m$ that are sufficiently small such that each contains at most one galaxy, and index these bins with $j$ (previously $\lambda i$). If $\mu_j$ is the Poisson mean for the apparent magnitude and redshift corresponding to $j$, $N_j$ is the number of galaxies in this bin, and $\mathbf{p}$ is a vector of all parameters, then the likelihood is given by:
\begin{equation}
\label{eq:genlike}
\mathcal{L}(\mathbf{p})=\prod_j\frac{\mu_j(\mathbf{p})^{N_j}e^{-\mu_j(\mathbf{p})}}{N_j!}.
\end{equation}
Using that the number of galaxies in bin $j$ is by construction equal to either $0$ or $1$, we can write the log-likelihood as:
\begin{eqnarray}
\nonumber
\ln{\mathcal{L}(\mathbf{p})}&=&\sum_j\left\{N_j\ln[\mu_j(\mathbf{p})]-\mu_j(\mathbf{p})-\ln[N_j!]\right\}\\
\label{eq:loglike}
&=&\sum_j\ln[\mu_j(\mathbf{p})]-\int\!\!\!\int\!\mu(\mathbf{p})\,\mathrm{d}m\,\mathrm{d}z.
\end{eqnarray}
The second term is a sum over all bins, regardless of whether there is a galaxy in that bin, and so can be replaced by an integral over all the probed redshifts and apparent magnitudes. The first term, on the other hand, is only non-zero for bins that contain a galaxy, and so can be viewed as a sum over all galaxies in the sample, rather than a sum over bins.

Similar to equation \eqref{eq:Nlambdai}, we can write:
\begin{eqnarray}
\label{eq:logliketerm2}
\int\!\!\!\int\!\mu(\mathbf{p})\,\mathrm{d}m\,\mathrm{d}z\!\!\!\!\!&=&\!\!\!\!\! B\int_0^{z_\mathrm{max}}\frac{d_\mathrm{c}(z)^2\,\phi_*(\mathbf{p},z)}{\sqrt{\Omega_\mathrm{m,0}(1+z)^3+\Omega_{\Lambda,0}}}\times\\
\nonumber
\!\!\!\!\!& &\!\!\!\!\! \left[\Gamma\left(\alpha(\mathbf{p},z)+1,10^{\frac{2}{5}(M_*(\mathbf{p},z)-\Mlim)}\right)\right]^{-1}\times\\
\nonumber
\!\!\!\!\!& &\!\!\!\!\! \left[\Gamma\left(\alpha(\mathbf{p},z)+1,10^{\frac{2}{5}(M_*(\mathbf{p},z)-M_\mathrm{max}(z))}\right)-\right.\\
\nonumber
\!\!\!\!\!& &\!\! \left.\Gamma\left(\alpha(\mathbf{p},z)+1,10^{\frac{2}{5}(M_*(\mathbf{p},z)-M_\mathrm{min}(z))}\right)\right]\,\mathrm{d}z,
\end{eqnarray}
where $M_\mathrm{max}(z)=\mathrm{min}[M(m_\mathrm{max},z),M_\mathrm{lim}]$ and $M_\mathrm{min}(z)=M(m_\mathrm{min},z)$. In our application, $M_\mathrm{lim}=-16$, $z_\mathrm{max}=0.8$, $m_\mathrm{min}=13$ and $m_\mathrm{max}=21$.

In applying this maximum-likelihood method (\S\ref{subsubsec:direct}), we assume full information on each galaxy, meaning we can use its absolute magnitude and redshift directly. For galaxy $j$, we can therefore write (see equation~\eqref{eq:Nlambdai_pre}):
\begin{eqnarray}
\nonumber
\ln[\mu_j(\mathbf{p})]\!\!\!\!\!&=&\!\!\!\!\!\ln\left[\frac{2}{5}\ln(10)B\right]+2\ln\left[d_\mathrm{c}(z_j)\right]+\ln\left[\phi_*(\mathbf{p},z_j)\right]-\\
\nonumber
\!\!\!\!\!& &\!\!\!\!\! \frac{1}{2}\ln\left[\Omega_\mathrm{m,0}(1+z_j)^3+\Omega_{\Lambda,0}\right]+\frac{2}{5}\ln(10)\left(\alpha(\mathbf{p},z_j)+1\right)\times\\
\nonumber
\!\!\!\!\!& &\!\!\!\!\! \left(M_*(\mathbf{p},z_j)-M_j\right)-10^{\frac{2}{5}\left(M_*(\mathbf{p},z_j)-M_j\right)}-\\
\label{eq:logliketerm1}
\!\!\!\!\!& &\!\!\!\!\! \ln\left[\Gamma\left(\alpha(\mathbf{p},zj)+1,10^{\frac{2}{5}\left(M_*(\mathbf{p},z_j)-M_\mathrm{lim}\right)}\right)\right].
\end{eqnarray}

\section{Ignored effects}
\label{app:ignored}
\begin{figure*}
\begin{center}
\begin{tabular}{ccc}
\includegraphics[width=1.0\columnwidth, trim=20mm -2mm 13mm 8mm]{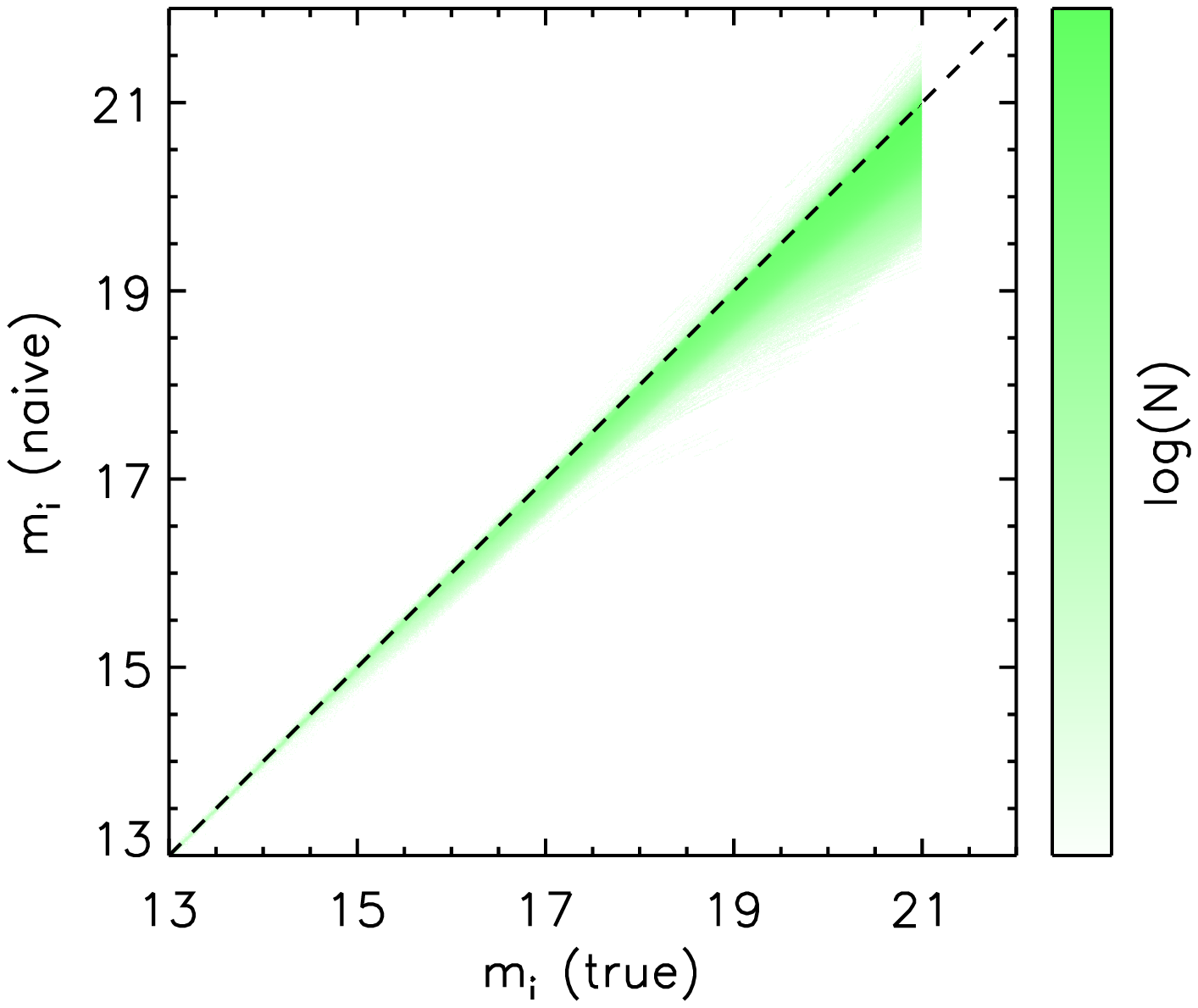} & &
\includegraphics[width=1.0\columnwidth, trim=30mm -2mm 3mm 8mm]{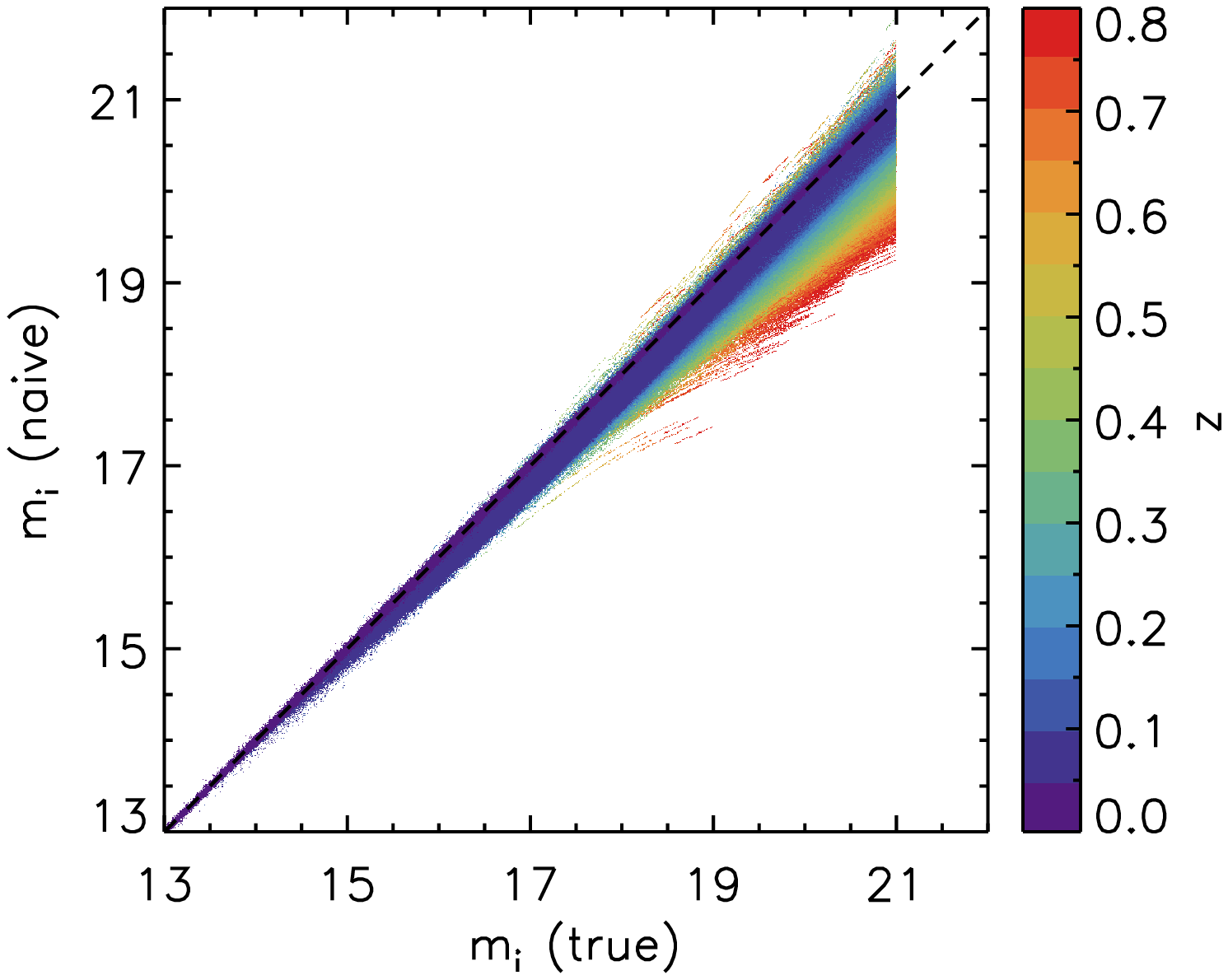}
\end{tabular}
\caption{A comparison of the apparent $i$-band magnitude naively derived from a galaxy's absolute magnitude, and its true value. The left-hand panel shows the log-density of galaxies in the plane, while the right-hand panel shows the distribution with redshift. Fainter galaxies and in particular high-redshift galaxies need larger K-corrections and are therefore more sensitive to these being ignored.}
\label{fig:magmag}
\end{center}
\end{figure*}

The most important effects that we ignore in our model are (i) the fact that galaxy spectra are not flat and that therefore the relation between apparent and absolute magnitude is not straightforward, and (ii) biasing due to lensing (de)magnification. Incorporating either effect into our model or even quantify how ignoring them impacts our results is far from trivial and outside the scope of this paper. However, in order to gauge the importance of the former effect -- that is, ignoring K-corrections -- we show in Figure~\ref{fig:magmag} a comparison between the true $i$-band apparent magnitude of each mock galaxy as calculated by \citet{Henriques2015} and the one naively derived from that galaxy's $i$-band absolute magnitude through $m_i=M_i-5\left[1-\log_{10}(d_\mathrm{L}(z))\right]$, with $d_\mathrm{L}(z)$ the luminosity distance. All galaxies in the light cone that satisfy the redshift and (true) magnitude cuts of our catalogue are included.

In the left-hand panel of Figure~\ref{fig:magmag}, brightness indicates the logarithmic density of galaxies at each point in the space. This shows that the effect of ignoring K-corrections is largest for the faintest galaxies, and that the naive relation tends to overestimate the apparent brightness of these galaxies. In the right-hand panel, we show now colour-code by redshift. As the redshift increases, the mean true apparent magnitude and the mean difference between it and the naive apparent magnitude increase as well. This shows that, as one might expect, the effect of ignoring K-corrections is strongest at high redshifts.

In our case (with $\Delta m=0.1$), taken over all redshifts the difference between the true and naive apparent magnitude is at most one bin for the majority of galaxies, but even this small effect may be enough to significantly impact the results when applying our current model to real data. Other photometric bands may be affected differently. More work is needed to explore and account for this.

\bsp
\label{lastpage}
\end{document}